\documentclass{article}

\usepackage[english]{babel}
\usepackage{jheppub}

\usepackage{amsmath}
\usepackage{graphicx}

\usepackage{booktabs}
\usepackage{multirow}
\usepackage{makecell}
\usepackage{array}
\usepackage{orcidlink}

\allowdisplaybreaks

\def\eps{\epsilon}

\preprint{TTP26-009, P3H-26-020, USTC-ICTS/PCFT-26-19}

\title{Two-loop Six-point Planar Massless Feynman Integrals
to Higher $\epsilon$ Orders}
\abstract{In this work, we calculate two-loop six-point planar massless Feynman integrals at higher orders in the dimensional regulator $\epsilon$, corresponding to higher transcendental weights. In previous works, these integrals were calculated up to weight four for the purpose of two-loop gauge theory amplitudes. Using modern rational reconstruction methods, we identify the complete alphabet with $269$ letters relevant to all weights, derive the analytic canonical differential equation and obtain the symbols up to weight six. As a proof of concept, using a new method with Chebyshev pseudospectral transport, we show that the corresponding pure basis can be efficiently evaluated up to weight six, i.e., to $  \mathcal{O}(\epsilon^2)$ in a physical scattering region. The results of this work can be applied to future three-loop amplitudes and provide new data for the formal study of symbols and cluster algebras.
}

\author[a]{Yuanche Liu,\,\orcidlink{0009-0008-4604-1306}}
\author[b]{Antonela Matija\v{s}i\'{c},\,\orcidlink{0000-0001-6477-6092}}
\author[c]{Tiziano Peraro\,\orcidlink{0000-0003-1534-4378}}
\author[d]{Yingxuan Xu\,\orcidlink{0000-0001-6135-8864}}
\author[e]{Zihua Yang\,\orcidlink{0009-0003-3697-1554}}
\author[a,f,g]{Yang Zhang\,\orcidlink{0000-0001-9151-8486}}

\affiliation[a]{Interdisciplinary Center for Theoretical Study,
University of Science and Technology of China,
Hefei, Anhui 230026, China}

\affiliation[b]{PRISMA Cluster of Excellence, Institut für Physik, Staudinger Weg 7,
Johannes Gutenberg-Universität Mainz, D - 55099 Mainz, Germany}

\affiliation[c]{Dipartimento di Fisica e Astronomia, Universit\'{a} di Bologna e INFN, Sezione di Bologna, via Irnerio 46, I-40126 Bologna, Italy}

\affiliation[d]{Institute for Theoretical Particle Physics, Karlsruhe Institute of Technology (KIT), Wolfgang-Gaede-Straße~1, 76131 Karlsruhe, Germany}

\affiliation[e]{School of Physical Sciences, University of Science and Technology of China,
Hefei, Anhui 230026, China}

\affiliation[f]{Peng Huanwu Center for Fundamental Theory,
Hefei, Anhui 230026, China}

\affiliation[g]{Center for High Energy Physics, Peking University,
Beijing 100871, People’s Republic of China}

\emailAdd{liuyuanche@mail.ustc.edu.cn}
\emailAdd{amatijas@uni-mainz.de}
\emailAdd{tiziano.peraro@unibo.it}
\emailAdd{yingxuan.xu@kit.edu}
\emailAdd{yangzihua@mail.ustc.edu.cn}
\emailAdd{yzhphy@ustc.edu.cn}

\begin{document}

\maketitle
\section{Introduction}

Over the past decade, remarkable progress has been achieved in the analytic computation of multi-loop, multi-leg scattering amplitudes. In particular, advances in the evaluation of two-loop Feynman integrals, together with the formulation of their results in terms of iterated integrals, have enabled precise predictions for increasingly complex collider processes. As a result, analytic results for two-loop five-point integrals and amplitudes are by now well established and have found widespread phenomenological applications~\cite{Gehrmann:2015bfy,Gehrmann:2018yef,Abreu:2018aqd,Chicherin:2018old,Chicherin:2020oor,Abreu:2021oya,Chawdhry:2021mkw,Agarwal:2021vdh,Czakon:2021mjy}. Recent work has pushed the multi-loop multi-leg frontier further in two complementary directions: major progress has been achieved for massless planar three-loop five-point Feynman integrals~\cite{Liu:2024ont,Chicherin:2025mvc,Chicherin:2025jej}; on the other hand, this progress has been extended to planar massless two-loop six-point scattering, where complete analytic solutions for the master integrals and their associated function spaces, up to the transcendental weight four, have been obtained~\cite{Henn:2021cyv,Henn:2024ngj,Abreu:2024fei,Henn:2025xrc}. The corresponding function space is a key input for the recent breakthrough in the bootstrap calculation of Wilson loop with Lagrangian insertion \cite{Carrolo:2025agz} and QCD amplitudes \cite{Carrolo:2025pue,Carrolo:2026qpu}.

These results on two-loop six-point integrals provide full control over the divergent and finite parts of two-loop six-point amplitudes and are sufficient for evaluating scattering amplitudes in four dimensions up to the finite part \cite{Abreu:2024fei,Henn:2025xrc}. However, this does not yet exhaust the information carried by the corresponding master integrals. In more general perturbative settings, two-loop six-point integrals will be used as lower-loop building blocks inside higher-order studies of infrared structure, and in such situations higher orders in the dimensional regulator $\epsilon$ become relevant.

Therefore, a primary motivation for going beyond the finite part comes from future analyses of planar three-loop six-point amplitudes. The infrared singularities of multi-leg amplitudes are controlled by universal factorization formulae, in which singular operators at higher loops multiply lower-loop building blocks~\cite{Catani:1998bh,Becher:2009qa}. As a result, when analyzing the infrared organization of three-loop six-point amplitudes, it is natural to expect that higher-order terms in the $\epsilon$-expansion of the two-loop six-point master integrals will be used. In particular, the weight-six part of these integrals is expected to enter as relevant input in such analyses. From this perspective, extending the two-loop six-point integrals beyond the finite part is  a natural step toward six-particle scattering at higher perturbative orders.

Beyond this practical motivation, access to higher orders in the $\epsilon$-expansion of the two-loop six-point integrals is also of intrinsic theoretical interest. Making these terms explicit gives direct access to the corresponding higher-weight symbol and function space. This is especially interesting in the six-point case, where cluster-algebraic structures have long been observed in the study of scattering amplitudes. Early work identified the role of cluster coordinates and cluster polylogarithms in planar amplitudes~\cite{Golden:2013xva,Golden:2014xqa}, while subsequent analyses at six and seven points clarified the organization of weight-four cluster functions and cluster adjacency properties~\cite{Harrington:2015bdt,Golden:2019kks}. Cluster-algebraic ideas were later extended to Feynman integrals and related function spaces~\cite{Chicherin:2020umh}. More recently, new studies have analyzed six-point symbol alphabets and singularity structures from the perspective of partial flag varieties and flag cluster algebras~\cite{Bossinger:2025rhf,Pokraka:2025ali}. Our results provide explicit higher-weight analytic data for the two-loop six-point master integrals themselves, and therefore offer information complementary to these structural analyses, beyond the previously known weight-four regime.

In this work, we extend the planar massless two-loop six-point master integrals to higher orders in the dimensional regulator $\epsilon$, i.e., with higher transcendental weights. With modern reconstruction methods, we analytically obtained all the previously not-yet-fitted off-diagonal entries \cite{Henn:2025xrc} of the differential equation for two-loop six-point integrals, which come from the weight-six part of these integrals. Six genuinely new letters are identified in this work. Together with the $18$ letters from the diagonal block of the top-sector differential equation of the double pentagon integral family~\cite{Henn:2021cyv}, there are $24$ more letters in addition to the $245$ letters in~\cite{Abreu:2024fei,Henn:2025xrc}
for the up-to-weight-four part. The $269$-letter alphabet is complete for two-loop six-point planar massless integrals to all orders in $\epsilon$.

From the differential equation, we  obtain the symbols for two-loop six-point planar massless integrals up to the transcendental weight six, i.e., $O(\epsilon^2)$. In principle, with enough RAM and computational time, it is possible to obtain the corresponding symbol to even higher order. With the symbol expression, we studied the symbol for the higher-weight parts of two-loop six-point planar massless integrals. This extends the available information beyond the previously known finite part and provides more complete input for future higher-loop applications.

To support this analysis, we also develop a new numerical approach for solving the canonical differential equations up to weight six, based on Chebyshev pseudospectral transport. The method provides a highly efficient and accurate way to transport boundary data and to evaluate the master integrals numerically. To the best of our knowledge, this is the first application of Chebyshev pseudospectral transport in this context. As a proof of concept, we present explicit numerical solutions, through $O(\epsilon^2)$, in a part of the physical scattering region, showing that the method provides a highly efficient and reliable tool for solving the canonical differential equations. A more complete exploration of the physical $2\to4$ scattering region is left for future work.

The paper is organized as follows. In Section~\ref{section:kinematics}, we introduce the kinematics and conventions. Section~\ref{section:CDE} presents the canonical differential equations, including the pure basis, and the reconstruction of the off-diagonal block. In Section~\ref{section:symbology}, we discuss the alphabet to all orders in $\epsilon$, the dihedral group action and the corresponding symbology. In Section~\ref{section:numerical evaluation}, we present the numerical evaluation of the differential equations, including the choice of kinematic regions and boundary values and the Chebyshev pseudospectral transport of the canonical system. We conclude in Section~\ref{section:summary} with a summary and outlook. Appendix~\ref{app:alphabet} reviews the known two-loop six-point alphabet up to weight-four order, while Appendix~\ref{app:OldToNewF4} explains the conversion between the previous 18-letter alphabet and the new alphabet.

The data of this work can be downloaded from 
\begin{quotation}
\url{https://bitbucket.org/yzhphy/2l6p_complete_function_space/src/main/version_2026/}\ ,
\end{quotation}
where a tutorial script is also included for readers to get familiar with two-loop six-point kinematics and integrals. 

\section{Kinematics and Conventions}
\label{section:kinematics}

We follow the kinematic conventions of Ref.~\cite{Henn:2021cyv}, as well as~\cite{Henn:2025xrc}. 
The six external momenta are denoted by $p_i$ with $i=1,\ldots,6$, and satisfy
\begin{equation}
  p_i^2 = 0\,, \qquad \sum_{i=1}^6 p_i = 0\,.
\end{equation}
The Mandelstam variables are
\begin{equation}
  \label{eq:2}
  s_{ij}=(p_i+p_j)^2\,,\quad s_{ijk} = (p_i + p_j + p_k)^2 \,,\quad 1\leq i,j,k\leq 6\,.
\end{equation}
Moreover, the notation for the Gram determinant reads
\begin{equation}
  \label{eq:3}
  G\left(
\begin{array}{ccc}
u_1 &\ldots & u_n\\
v_1 & \ldots & v_n
\end{array}\right) = {\rm det} (2u_i \cdot v_j)\,,
\end{equation}
where the right-hand side denotes the determinant of the $n\times n$ matrix with entries $(2\,u_i\!\cdot v_j)$, $1\leq i,j\leq n$. Note that the Gram matrix normalization is different from that in~\cite{Henn:2021cyv}. Thus, the related formulae in~\cite{Henn:2021cyv} are  adopted with the new normalization in this work. For convenience, we introduce the shorthand
\begin{equation}
  G(i_1,\ldots,i_k) \equiv 
  G\!\left(
  \begin{array}{ccc}
  p_{i_1} & \ldots & p_{i_k}\\
  p_{i_1} & \ldots & p_{i_k}
  \end{array}\right)\!,
  \qquad 1\leq i_1,\ldots,i_k\leq 6\,.
\end{equation}

We work in a scheme such that all external momenta are in four-dimensional Minkowski space~\cite{Henn:2021cyv,Abreu:2024fei,Henn:2025xrc}. For other schemes such that external momenta are  $D$-dimensional, there are nine independent Mandelstam invariants~\cite{eden2002analytic}. In our scheme, however, any set of five momenta is linearly dependent, which implies the constraint
\begin{equation}
  G(1,2,3,4,5)=0\,.
  \label{eq:GramConstraint}
\end{equation}
Consequently, only eight Mandelstam invariants are independent in four dimensions.

In addition to scalar invariants, the pseudo scalars are defined as
\begin{equation}
  \epsilon_{ijkl} \equiv 
  4\,\mathrm{i}\,\varepsilon_{\mu_1\mu_2\mu_3\mu_4}\,
  p_i^{\mu_1} p_j^{\mu_2} p_k^{\mu_3} p_l^{\mu_4}\,,
  \qquad 1\leq i,j,k,l\leq 6\,,
\end{equation}
where $\varepsilon_{\mu_1\mu_2\mu_3\mu_4}$ is the four-dimensional Levi-Civita tensor. It follows that
\begin{equation}
    \epsilon_{ijkl} ^2=G(i,j,k,l)\,.
\end{equation}

With the convention $p_\mu \sigma^{\mu,\alpha\dot\alpha} \equiv \lambda^\alpha \tilde \lambda^{\dot \alpha}$, the spinor products are defined as,
\begin{equation}
    \langle ij \rangle \equiv \lambda_i^\alpha \lambda_{j,\alpha}, \qquad [ij] \equiv \tilde{\lambda}_{i,\alpha} \tilde{\lambda}_j^\alpha
\end{equation}
with $s_{ij}=\langle ij\rangle [ji]$. It is then very useful to introduce {\it momentum twistor variables} for the kinematics
\begin{equation}
    Z_i \equiv \begin{pmatrix} \lambda_{i\alpha} \\ \mu_{i,\dot{\beta}} \end{pmatrix},\quad i=1,\ldots 6\,.
\end{equation}
The $4\times 6=24$ components of the momentum twistors $Z_i$ can be parametrized in terms of eight independent variables. Such a parametrization is called momentum twistor parametrization \cite{Hodges:2009hk}. Various choices of parametrization are given in \cite{Bourjaily:2010wh,Badger:2013gxa, Henn:2021cyv,Henn:2024ngj,Abreu:2024fei}. With momentum twistor variables, all pseudo scalars and the spinor products are rationalized. Additionally, the constraint~\eqref{eq:GramConstraint} is automatically satisfied which enables us to work with eight independent variables.

As shown in Ref.~\cite{Henn:2021cyv}, the genuine planar two-loop six-point topologies can be classified into the double-pentagon (DP) and hexagon-box (HB) families. For these DP and HB families, we introduce the following inverse propagators:
\begin{gather}
D_1=l_1^2,D_2=\left(l_1-p_1\right){}^2,D_3=\left(l_1-p_1-p_2\right){}^2,D_4=\left(l_1-p_1-p_2-p_3\right){}^2,\nonumber
\\
D_5=\left(l_1+l_2\right){}^2,D_6=l_2^2,D_7=\left(l_2+p_1+p_2+p_3+p_4+p_5\right){}^2,D_8=\left(l_2+p_1+p_2+p_3+p_4\right){}^2,\nonumber
\\D_9=\left(l_2+p_1+p_2+p_3\right){}^2,D_{10}=\left(l_1+p_5+p_6\right){}^2,D_{11}=\left(l_2+p_1+p_2\right){}^2 \,.
\label{eq:Propagators}
\end{gather}
This allows us to treat all integrals in DP and HB families using one unified notation.
For example, the DP family corresponds to the sector $(1,1,1,1,1,1,1,1,1,0,0)$,  the HB family corresponds to $(1,1,1,1,1,1,1,1,0,1,0)$.

For the study of planar Feynman integrals, it is important to consider the corresponding dihedral group. The dihedral group $D_6$, which is a subgroup of the permutation group $S_6$, can be generated by the two permutations
\begin{align}
    \mathcal T:& 1\to 2, 2\to 3, 3\to 4, 4\to 5, 5\to 6, 6\to 1\,,\\
    \mathcal R:& 1\to 6, 2\to 5, 3\to 4, 4\to 3, 5\to 2, 6\to 1
    \label{eq:dihedral-action}
\end{align}
with the rules $\mathcal T^6=\mathcal I$, $\mathcal R^2=\mathcal I$ and $\mathcal R \mathcal T \mathcal R=\mathcal T^{-1}$.

\section{Canonical Differential Equations}
\label{section:CDE}
In the recent papers~\cite{Abreu:2024fei} and~\cite{Henn:2025xrc}, two-loop six-point planar massless integrals were calculated up to transcendental weight four. More specifically, the corresponding pure integral basis, with the normalization such that the $\epsilon^{-4}$ order has weight zero, is calculated to $\epsilon^0$. It is shown in~\cite{Abreu:2024fei} and~\cite{Henn:2025xrc} that these results are sufficient for constructing two-loop amplitudes for gauge theories in four dimensions to $\epsilon^0$.

However, in the future, for the infrared subtraction of three-loop six-point massless amplitudes, higher orders in $\epsilon$, i.e., the parts of two-loop six-point pure integrals with transcendental weight higher than four, will be needed. This motivates our computation of two-loop six-point massless integrals to higher orders in $\epsilon$. Furthermore, as we will see, this higher-order computation leads to new letters for two-loop six-point integrals. Theoretically, our work thus provides interesting new data for the formal study of symbology and cluster algebras.

In this section, we review the results obtained in~\cite{Abreu:2024fei,Henn:2025xrc} and summarize the strategies used to reconstruct the off-diagonal entries of the differential equation for the DP family.

\subsection{Review of the pure integral basis}
Recall that in~\cite{Abreu:2024fei} and~\cite{Henn:2025xrc}, up to the transcendental weight four, the two-loop six-point planar massless pure integrals contain $245$ letters, which are closed under the dihedral group action. For the HB family, these $245$ letters are sufficient to reconstruct the corresponding canonical differential equation. For the DP family, these letters are sufficient to reconstruct the rows in the canonical differential equation for the sub-sector integrals with fewer-than-nine propagators.

For the top sector of the DP family with nine propagators, Ref.~\cite{Abreu:2024fei} proves that the top-sector integrals up to the weight four can be reduced to linear combinations of lower-sector integrals and evanescent integrals (vanishing up to the order $O(\epsilon)$). In~\cite{Henn:2025xrc}, the pure integral candidates for the DP top sector from~\cite{Henn:2021cyv} were confirmed from the numerical canonical differential equation. Furthermore, those top-sector pure integrals, up to the weight four, are explicitly reduced to linear combinations of lower-sector integrals. For these reasons, the top sector integrals do not provide independent functions up to weight four, so the rows in the canonical differential equation for the DP top sector were not analytically reconstructed in~\cite{Abreu:2024fei} and~\cite{Henn:2025xrc}. Possible new letters from these rows, for the higher-order parts in $\epsilon$ of the DP top-sector integrals, were not identified either.

In this work, we first identify the complete alphabet for two-loop six-point planar massless integrals to all orders in $\epsilon$. To achieve this, it is sufficient to determine the new letters for the rows corresponding to the DP top-sector integrals in the canonical differential equation. Recall that the DP family contains a pure integral basis with $267$ integrals in total~\cite{Henn:2025xrc}, namely $I_i^{\text{DP}}$, $i=1,\ldots, 267$. The top sector contains five pure integrals, which are denoted as $I_i^{\text{DP}}$, $i=1,\ldots, 5$, with the following integrand form~\cite{Henn:2021cyv}:
\begin{align}
I_1^{\text{DP}} &= \int \frac{d^{4-2\epsilon} l_1}{i\pi^{2-\epsilon}} \frac{d^{4-2\epsilon} l_2}{i\pi^{2-\epsilon}} \frac{N_1^{\text{DP-a}} - N_4^{\text{DP}}}{D_1 \dots D_9}, \\
I_2^{\text{DP}} &= \int \frac{d^{4-2\epsilon} l_1}{i\pi^{2-\epsilon}} \frac{d^{4-2\epsilon} l_2}{i\pi^{2-\epsilon}} \frac{N_2^{\text{DP}} - N_3^{\text{DP}}}{D_1 \dots D_9}, \\
I_3^{\text{DP}} &= \frac{ s_{345} \Delta_6}{8G(1,2,3,6)} \int \frac{d^{4-2\epsilon} l_1}{i\pi^{2-\epsilon}} \frac{d^{4-2\epsilon} l_2}{i\pi^{2-\epsilon}} \frac{G\begin{pmatrix} l_1 & p_1 & p_2 & p_3 & p_6 \\ l_2 & p_1 & p_2 & p_3 & p_6 \end{pmatrix}}{D_1 \dots D_9}, \\
I_4^{\text{DP}} &= F_4 \epsilon^2 \int \frac{d^{6-2\epsilon} l_1}{i\pi^{3-\epsilon}} \frac{d^{6-2\epsilon} l_2}{i\pi^{3-\epsilon}} \frac{1}{D_1 \dots D_9}, \\
I_5^{\text{DP}} &= \int \frac{d^{4-2\epsilon} l_1}{i\pi^{2-\epsilon}} \frac{d^{4-2\epsilon} l_2}{i\pi^{2-\epsilon}}\left[ \frac{N_1^{\text{DP}} + N_4^{\text{DP}} }{D_1 \dots D_9} + \frac{F_5}{4 G(1,2,3,6)} \frac{G\begin{pmatrix} l_1 & p_1 & p_2 & p_3 & p_6 \\ l_2 & p_1 & p_2 & p_3 & p_6 \end{pmatrix}}{D_1 \dots D_9} \right], 
\end{align}
where
\begin{align}
    \Delta_6 &= \langle 12\rangle [23] \langle 34 \rangle [45] \langle 56 \rangle [61]-[12]\langle 23 \rangle [34] \langle  45\rangle [56] \langle 61\rangle ,\\
    F_4 & =\frac{1}{8} \Bigg( G(1,2,3,6)s_{45}^2 + G(1,2,3,5)s_{46}^2 + G(1,2,3,4)s_{56}^2 \notag \\ 
&- 2\epsilon_{1235}\epsilon_{1236}s_{45}s_{46} - 2\epsilon_{1234}\epsilon_{1236}s_{45}s_{56} - 2\epsilon_{1234}\epsilon_{1235}s_{46}s_{56} \Bigg)^{1/2} \\
&=\frac{1}{8}\sqrt{\lambda(s_{45}\eps_{1236},s_{46} \eps_{1235},s_{56} \eps_{1234})},
\label{eq:F4}
\end{align}
and $\lambda$ denotes the K\"all\'en function
\begin{equation}
    \lambda(a,b,c)=a^2+b^2+c^2-2ab-2ac-2bc.
    \label{eq:Kallen}
\end{equation}
We emphasize that $F_4$ has a simple form in terms of the K\"all\'en function. The explicit expressions for the constant $F_5$ and the chiral numerators $N_i^\text{DP}$, $i=1,\ldots,4$ in terms of momentum twistor variables, are given in~\cite{Henn:2021cyv}.

Note that $I^\text{DP}_1$ and $I^\text{DP}_3$ are nonzero only starting at $\mathcal O(\epsilon^1)$, i.e. weight five, due to the dual conformal symmetry and evanescent Gram determinant numerator respectively. The first nonzero component of integral $I^\text{DP}_4$ starts at $O(\epsilon^2)$, i.e. weight six, because of the absence of infrared divergence. The nonzero parts of the remaining two integrals, $I^\text{DP}_2$ and $I^\text{DP}_5$, start at $\mathcal O(\epsilon^0)$, i.e. weight four, which means that they are finite.

In Ref.~\cite{Henn:2025xrc}, the pure integral basis for the DP family is confirmed by the $\epsilon$ factorization property in the differential equation \cite{Henn:2013pwa} calculated on numerical points 
\begin{equation}
    \mathrm d I_i^\text{DP} =\epsilon \sum_{j=1}^{267} \bigg(\mathrm d\tilde A_{ij} \bigg)I_j^\text{DP}.
\end{equation}
Rows $6$ to $267$ of the matrix $\tilde A_{ij}$ were analytically reconstructed in~\cite{Abreu:2024fei, Henn:2025xrc}, in terms of the logarithm of $245$ letters defined in~\cite{Abreu:2024fei, Henn:2025xrc}, respectively. In~\cite{Henn:2021cyv}, the top $5\times 5$ diagonal block, $\tilde A_{ij}$, $1\leq i\leq 5$,  $1\leq j\leq 5$, was also analytically reconstructed. Furthermore, as pointed out in~\cite{Henn:2025xrc}, the $245$ letters from~\cite{Abreu:2024fei, Henn:2025xrc} can also be used to fit analytically the first, second, third and the fifth rows, but {\it not the fourth} row of the off-diagonal block $\tilde A_{4j}$, $6\leq j\leq 267$. 

\begin{figure}
\includegraphics[scale=0.5]{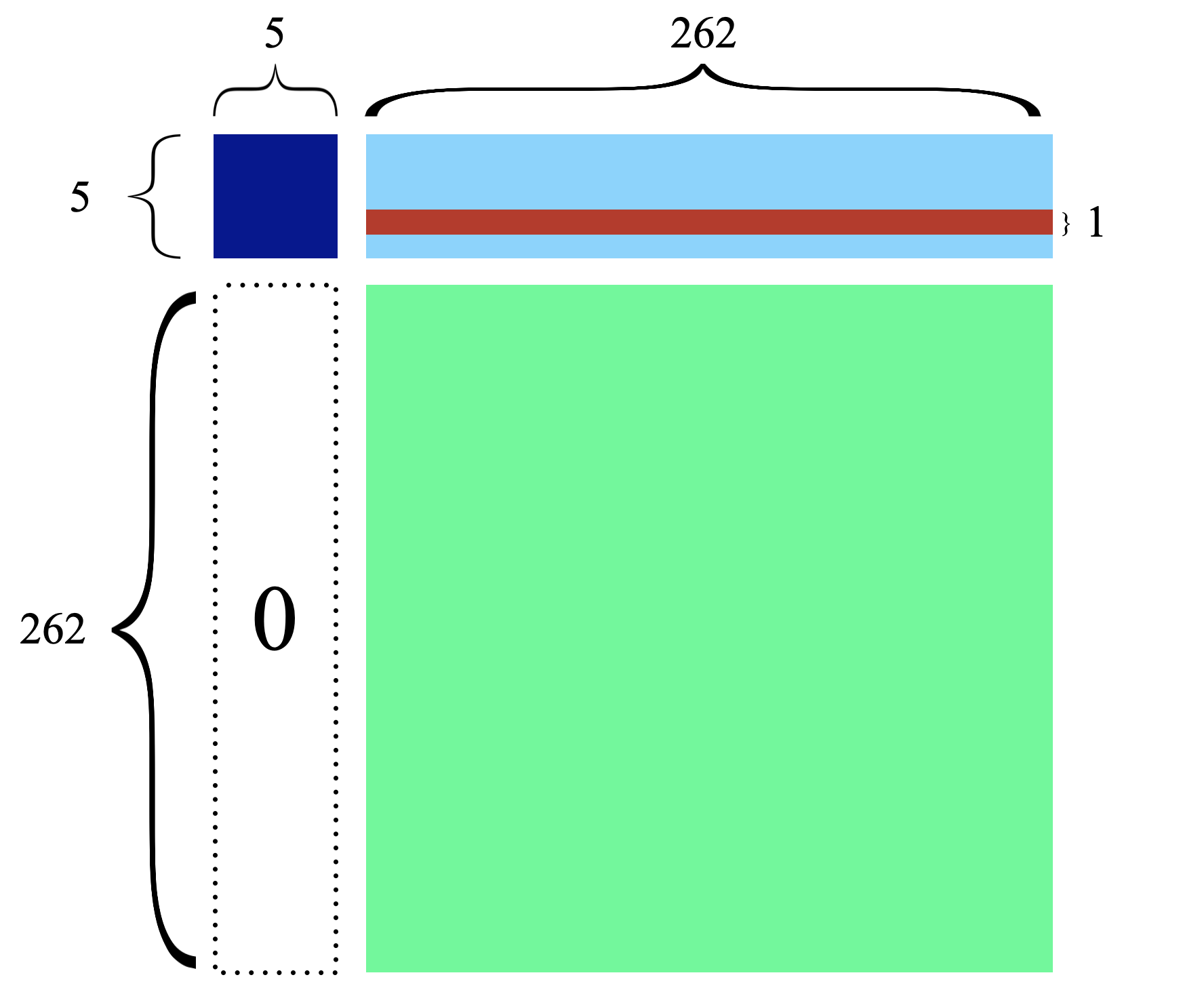}
\caption{The block structure of the differential equation for the DP family. The light green block represents the differential equation for the integrals with fewer than $9$ propagators, which were analytically reconstructed in~\cite{Abreu:2024fei} and~\cite{Henn:2025xrc}. Furthermore, the light blue area represents the first, second, third and fifth row of the off-diagonal block, which were analytically reconstructed in~\cite{Henn:2025xrc}. The deep blue block represents the cut differential equation for the top-sector integrals with $9$ propagators, which were reconstructed in~\cite{Henn:2021cyv}. In this work, we analytically reconstruct the fourth row of the off-diagonal part, represented by the carnelian red block. The numbers in the figure show the corresponding block sizes.}
\label{figure:DE_DP_Blocks}
\end{figure}

Thus, in this work, we analytically reconstruct the fourth row  of the differential equation for the DP family, to get the complete differential equation and the alphabet for two-loop six-point planar massless Feynman integrals to {\it all} $\epsilon$ orders. Fig.~\ref{figure:DE_DP_Blocks} illustrates the block structure of the differential equation for the DP family, emphasizing the new analytic computation in this work.

\subsection{Reconstruction of the off-diagonal block}

The goal is to analytically reconstruct the off-diagonal block to get the complete and analytic differential equation of the DP family. To achieve this, we used a combination of modern reconstruction strategies, which we summarize below:
\begin{enumerate}
    \item We first obtain the complete differential equation at a number of numerical kinematic points. Momentum twistor parametrization is used here to rationalize $\epsilon_{ijkl}$ and $\Delta_6$. The derivatives of the pure integrals in the eight momentum twistor variables are calculated, and then numerically reduced by standard IBP solvers like \texttt{FIRE7}~\cite{Smirnov:2025prc} or \texttt{KIRA3}~\cite{Lange:2025ofh} to produce the  differential equation at a numerical point. Typically, on a workstation with $20$ cores and $128$GB RAM, it takes $2\sim 4$ hours to get a numerical differential equation for the DP family in eight momentum twistor variables.  
    
    \item It is well known that before the reconstruction, it is  helpful to identify the linear relations between the matrix entries for the reconstruction~\cite{Badger:2021imn, Liu:2023cgs}. An elementary linear algebra exercise shows that, out of the fourth row's entries of the off-diagonal matrix, only three entries are linearly independent over $\mathbb Q$. 
    
    \item We use $245$ letters from~\cite{Abreu:2024fei, Henn:2025xrc} as well as the letters from the diagonal block in~\cite{Henn:2021cyv} to fit the three entries in the previous step. Out of the three entries, one can be analytically reconstructed in this way, but the two entries at positions $(4,6)$ and $(4,23)$ cannot. The entry located at $(4,23)$ is related to the entry $(4,6)$ by the reflection transformation $\mathcal R$. The former one, which corresponds to the entry $(4,6)$ of the matrix, would be the hard nut to crack.
    
    \item The last entry to be computed is the off-diagonal one between the pure integral $I_4^\text{DP}$ and an eight-propagator pure integral. This one contains a new letter. In order to fit it analytically in the momentum twistor variables, a significantly larger number of numerical points are needed. Traditional numerical IBP reduction in this situation becomes inconvenient. 
    We apply an octa-cut on the top sector which yields the off-diagonal block containing the new letter. The cut  increases the efficiency of numeric IBP reduction significantly, since it sets to zero most of the sectors.  For the reduction, we use the Baikov representation, where the integration variables are the $11$ generalized denominators, and a cut is achieved simply by setting the cut denominators to zero.  The reduction identities and the derivative operators with respect to the momentum twistor variables were then found, directly on the cut, using the \texttt{CALICO} package~\cite{Bertolini:2025zud}. The reduction identities were thus solved with \texttt{FiniteFlow}~\cite{Peraro:2019svx}, which was also used to analytically reconstruct the matrix elements of the differential equation which depend on the new letter, using the algorithms in~\cite{Peraro:2016wsq,Peraro:2019svx}.
    
    \end{enumerate}
After these steps, we obtain the complete analytic differential equation for the DP family, in terms of momentum twistor variables.

\section{The Alphabet to All Orders in $\eps$ and Symbology}
\label{section:symbology}
Starting from the complete analytic differential equation for the DP family, we are able to determine the full alphabet of double pentagon integrals to all orders in $\epsilon$.

In this section, we provide the complete alphabet required to represent planar two-loop six-point integrals at arbitrary order in $\eps$. We also describe how the dihedral group acts on the new letters and discuss the implications for the symbology.

\subsection{Alphabet}
All the letters beyond the $245$ letters in~\cite{Abreu:2024fei,Henn:2025xrc}, are 
associated with the pure integral $I_4^\text{DP}$, which is proportional to a square root $F_4$ \eqref{eq:F4}. Therefore, all those letters would be associated with $F_4$. $18$ such letters from the top $5\times 5$ diagonal block have been found in~\cite{Henn:2021cyv}, with the form
\begin{equation}
F_4, \quad \frac{F_4+R_i}{F_4-R_i}, \quad i=1,\ldots,5,
\end{equation}
and their cyclic permutations. $R_i$'s are rational functions in momentum twistor variables.

The ansatz for a new letter in the $(4,6)$ entry of the differential equation of DP family is
\begin{equation}
    \frac{-F_4+Q_6}{F_4+Q_6}\,.
\label{eq:new_letter}
\end{equation}
(The names $Q_i$, $i=1,\ldots 5$ are reserved for later use). The $d\log$ of this ansatz reads
\begin{equation}
\frac{2 Q_6 d F_4-2 F_4 d Q_6}{F_4^2-Q_6^2}\,.
\label{eq:dlog_Q6_odd_letter}
\end{equation}
We equate~\eqref{eq:dlog_Q6_odd_letter} with the analytic differential equation's $(4,6)$ entry and solve for $Q_6$. The result $Q_6$ is an analytic rational function in momentum twistor variables. 

The dihedral group $D_6$ acts on the analytic expression of the new letter~\eqref{eq:new_letter} to generate six independent letters in total. Recall that the other not-yet-fitted entry $(4,23)$ is related to this new letter by the reflection.  Therefore, we conclude that for all two-loop six-point planar massless integrals, there are 
 \begin{equation}
     245 + 18 + 6 = 269
 \end{equation}
independent letters. The linear independence is checked via the finite field packages \texttt{FiniteFlow}~\cite{Peraro:2019svx} and \texttt{Spasm}~\cite{spasm}. The complete alphabet is one of the key results of this work, which can be downloaded from
\begin{quotation}
\url{https://bitbucket.org/yzhphy/2l6p_complete_function_space/src/main/version_2026/alphabet/Alphabet.txt}\ .
\end{quotation}

The first $245$ letters for the pure integrals up to weight four are defined in~\cite{Henn:2025xrc}, and we recall their definitions in the Appendix~\ref{app:alphabet}. Furthermore, the $18$ letters given explicitly in~\cite{Henn:2021cyv} are rewritten in a more convenient form.

\begin{figure}
\includegraphics[scale=0.4]{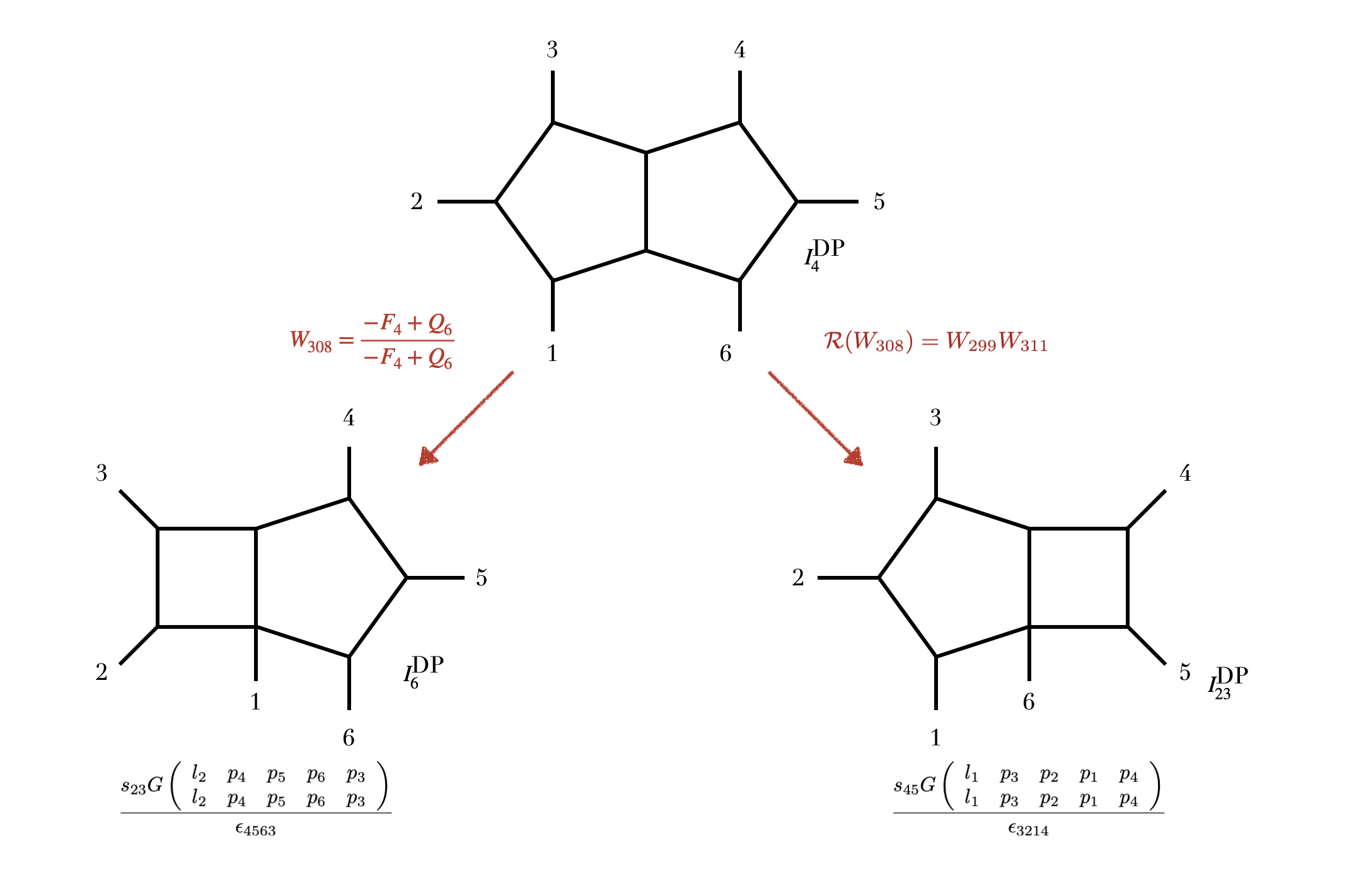}
\caption{The origin of the genuinely new letters $W_{308}$ and $W_{311}$. The differential equation matrix entry $(4,6)$ is the coefficient of derivative of $I_4^\text{DP}$ over $I_6^\text{DP}$, which provides a new letter $W_{308}$. Note that $I_4^\text{DP}$ is reflection invariant and the coefficient on reflected integral of $I_6^\text{DP}$, $I_{23}^\text{DP}$,  should be $d\log(\mathcal R(W_{308}))$. It is determined that $\mathcal R(W_{308}) =W_{299} W_{311}$, where $W_{311}$ is also a  new letter. The numerators for the pure integrals  $I_6^\text{DP}$ and $I_{23}^\text{DP}$ are listed.}
\label{figure:off_diagonal}
\end{figure}

In this work, the $18+6$ letters are chosen to be
\begin{align}
W_{290}&=F_4,\quad W_{291}=\mathcal T(W_{290}),\quad W_{292}=\mathcal T^2(W_{290}) \,, \label{eq:NewF4even}\\
W_{293}&=\frac{-F_4+Q_1}{F_4+Q_1},\quad 
 W_{293+i}=\mathcal T^i(W_{293}),\quad i=1,2,\\
 W_{296}&=\frac{-F_4+Q_2}{F_4+Q_2},\quad 
 W_{296+i}=\mathcal T^i(W_{296}),\quad i=1,2,\\
 W_{299}&=\frac{-F_4+Q_3}{F_4+Q_3},\quad 
 W_{299+i}=\mathcal T^i(W_{299}),\quad i=1,2,\\
 W_{302}&=\frac{-F_4+Q_4}{F_4+Q_4},\quad 
 W_{302+i}=\mathcal T^i(W_{302}),\quad i=1,2,\\
 W_{305}&=\frac{-F_4+Q_5}{F_4+Q_5},\quad 
 W_{305+i}=\mathcal T^i(W_{305}),\quad i=1,2,\\
 W_{308}&=\frac{-F_4+Q_6}{F_4+Q_6},\quad 
 W_{308+i}=\mathcal T^i(W_{308}),\quad i=1,\ldots, 5,
 \label{eq:NewF4odd}
\end{align}
where $Q_i$'s, $i=1,\ldots, 6$ are rational functions in momentum twistor variables. The letters $W_{290} ,\ldots, W_{307}$ span the same letter space as the 188th$\sim$205th letters defined in Ref.~\cite{Henn:2021cyv}. The conversion between the two choices is given in the Appendix~\ref{app:OldToNewF4}. $W_{308},\ldots, W_{313}$ are the genuinely new letters found in this work. The origin of the letters $W_{308}$ and $W_{311}$ is shown in Fig.~\ref{figure:off_diagonal}.

Since $F_4$ has the symmetry $\mathcal T^3(F_4)=F_4$, the $24$ letters associated with $F_4$ can also be classified into three sets: the first set reads
\begin{equation}
    F_4,\quad \frac{-F_4+Q_1}{F_4+Q_1},\quad  \ldots ,\quad \frac{-F_4+Q_6}{F_4+Q_6},\quad \frac{-F_4+\mathcal T^3(Q_6)}{F_4+\mathcal T^3(Q_6)},
\end{equation}
and the other two sets are generated by the action $\mathcal T$ and $\mathcal T^2$ on the first set. $7$ odd letters are associated with each of the three roots, $F_4$, $\mathcal T (F_4)$ and $\mathcal T^2 (F_4)$.

Note that the logarithms of the letters $W_{293}\sim W_{313}$ change sign under the Galois transformation $F_4 \to -F_4$, so they are called odd letters associated with $F_4$. An interesting phenomenon is the factorization property of odd letters discussed in  Ref.~\cite{Heller:2019gkq}. It is expected that an odd letter like $(-F_4+Q)/(F_4+Q)$ factorizes as
\begin{equation}
    F_4^2-Q^2 = c \prod_j  {W_j}^{e_j}
\end{equation}
where $c$ is a rational number, $W_j$'s are even letters under this Galois transformation and $e_j$'s are integers. This factorization property can be used to determine the polynomials $Q$ from the knowledge of the even alphabet and the definition of a square root~\cite{Matijasic:2024too}. We used a modified version of \verb|Effortless|~\cite{Effortless} to explicitly construct polynomials $Q_i$ such that this factorization is satisfied, for instance:
\begin{align}
 F_4^2-Q_1^2&=-\frac{1}{16} W_7 W_{31} W_{123} W_{128} =-\frac{1}{16}s_{123} s_{46} \epsilon_{1234}\epsilon_{6123}\,,\\
F_4^2-Q_2^2&=-\frac{1}{16} W_1 W_2 W_4 W_5 W_{28} W_{31} =-\frac{1}{16}s_{12}s_{23} s_{13}s_{45} s_{46}s_{56}\,,\\
F_4^2-Q_3^2&=\frac{1}{16} W_4 W_5 W_{123} W_{128}=\frac{1}{16} s_{45}s_{56}\epsilon_{1234}\epsilon_{6123}\,,\\
F_4^2-Q_4^2&=\frac{1}{16}\frac{W_1 W_2 W_7 W_{28} W_{126}^2}{W_{23}^2}=\frac{1}{16}\frac{s_{12} s_{13} s_{23} s_{123}\epsilon_{4561}^2 }{(s_{12} + s_{13})^2}\,,\\
F_4^2-Q_5^2&=\frac{1}{16}\frac{W_1 W_2 W_7 W_{28} W_{125}^2}{W_{16}^2}=\frac{1}{16}\frac{s_{12} s_{13} s_{23} s_{123}\epsilon_{3456}^2 }{(s_{13} + s_{23})^2}\,,\\
F_4^2-Q_6^2&=\frac{1}{16} W_1 W_{28} W_{125} W_{132}=\frac{1}{16} s_{12} s_{13} \epsilon_{3456} \epsilon_{4562} \,.
\end{align}

Recently, the six-point symbol alphabet has been analyzed from the perspective of partial flag varieties and flag cluster algebras~\cite{Bossinger:2025rhf,Pokraka:2025ali}. With complete two-loop six-point alphabet at hand, it would be interesting to see whether the new letters can be described in the same manner. 
Additionally, during the preparation stage of this manuscript, a paper on cluster-algebraic letters for five- and six-point QCD processes \cite{Aliaj:2026iny} has appeared. The authors predict new letters for six-point scattering in QCD by breaking dual conformal invariance of nine-point amplitudes in $\mathcal{N}=4$ super Yang-Mills theory. However, it appears that the genuinely new letters in this work are not included in Ref.~\cite{Aliaj:2026iny}. 

The complete alphabet for two-loop six-point integrals to all orders in $\eps$ consisting of these $269$ letters, is denoted by $\mathbb A$. In other words, the $\tilde A$ matrix for the canonical differential equation for the double pentagon family reads
\begin{equation}
    \tilde A=\sum_{W \in \mathbb A} \alpha_W \log(W),
\end{equation}
where each $\alpha_W$ is a $267\times 267$ constant matrix. The analytic $\tilde A$ matrix is given at the following link:
\begin{quotation}
\url{https://bitbucket.org/yzhphy/2l6p_complete_function_space/src/main/version_2026/DE/}
\end{quotation}

\subsection{Dihedral group action}
The alphabet $\mathbb A$, consisting of $269$ letters, is closed under the action of the dihedral group $D_6$.  The dihedral group action on the $245$ letters is given in Ref.~\cite{Henn:2025xrc}, so here we give the actions of the letters associated with $F_4$ only.

Under the translation $\mathcal T$, the letters transform as:
\begin{equation}
\begin{aligned}
& \log(W_{290}) \to \log(W_{291}), \quad
\log(W_{291}) \to \log(W_{292}), \quad
\log(W_{292}) \to \log(W_{290}), \\
& \log(W_{293}) \to \log(W_{294}), \quad
\log(W_{294}) \to \log(W_{295}), \quad
\log(W_{295}) \to -\frac{1}{2} \log(W_{302}) - \frac{\log(W_{305})}{2}, \\
& \log(W_{296}) \to \log(W_{297}), \quad
\log(W_{297}) \to \log(W_{298}), \quad
\log(W_{298}) \to \log(W_{296}), \\
& \log(W_{299}) \to \log(W_{300}), \quad
\log(W_{300}) \to \log(W_{301}), \quad
\log(W_{301}) \to \frac{\log(W_{302})}{2} - \frac{\log(W_{305})}{2}, \\
& \log(W_{302}) \to \log(W_{303}), \quad
\log(W_{303}) \to \log(W_{304}), \quad
\log(W_{304}) \to \log(W_{299}) - \log(W_{293}), \\
& \log(W_{305}) \to \log(W_{306}), \quad
\log(W_{306}) \to \log(W_{307}), \quad
\log(W_{307}) \to -\log(W_{293}) - \log(W_{299}), \\
& \log(W_{308}) \to \log(W_{309}), \quad
\log(W_{309}) \to \log(W_{310}), \quad
\log(W_{310}) \to \log(W_{311}), \\
& \log(W_{311}) \to \log(W_{312}), \quad
\log(W_{312}) \to \log(W_{313}), \quad
\log(W_{313}) \to \log(W_{308})\,,
\end{aligned}
\end{equation}
while under the reflection $\mathcal R$, the letters transform as:
\begin{equation}
\begin{aligned}
& \log(W_{290}) \to \log(W_{290}), \quad
\log(W_{291}) \to \log(W_{292}), \quad
\log(W_{292}) \to \log(W_{291}), \\
& \log(W_{293}) \to -\frac{1}{2} \log(W_{302}) - \frac{\log(W_{305})}{2}, \quad
\log(W_{294}) \to \log(W_{295}), \quad
\log(W_{295}) \to \log(W_{294}), \\
& \log(W_{296}) \to \log(W_{296}), \quad
\log(W_{297}) \to \log(W_{298}), \quad
\log(W_{298}) \to \log(W_{297}), \\
& \log(W_{299}) \to \frac{\log(W_{305})}{2} - \frac{\log(W_{302})}{2}, \quad
\log(W_{300}) \to -\log(W_{301}), \quad
\log(W_{301}) \to -\log(W_{300}), \\
& \log(W_{302}) \to -\log(W_{293}) - \log(W_{299}), \quad
\log(W_{303}) \to \log(W_{307}), \quad
\log(W_{304}) \to \log(W_{306}), \\
& \log(W_{305}) \to \log(W_{299}) - \log(W_{293}), \quad
\log(W_{306}) \to \log(W_{304}), \quad
\log(W_{307}) \to \log(W_{303}), \\
& \log(W_{308}) \to \log(W_{299}) + \log(W_{311}), \quad
\log(W_{309}) \to \frac{\log(W_{304})}{2} - \frac{\log(W_{307})}{2} + \log(W_{310}), \\
&\log(W_{310}) \to \frac{\log(W_{303})}{2} - \frac{\log(W_{306})}{2} + \log(W_{309}),  \log(W_{311}) \to \frac{\log(W_{302})}{2} - \frac{\log(W_{305})}{2} + \log(W_{308}), \\
&\log(W_{312}) \to \log(W_{301}) + \log(W_{313}), \quad
\log(W_{313}) \to \log(W_{300}) + \log(W_{312})\,.
\end{aligned}
\end{equation}
Thus the closure of the alphabet $\mathbb A$ under the $D_6$ group action is evident.

We further explore the $D_6$ action on these $24$ letters associated with $F_4$. Recall that the dihedral group $D_6$ has six different representations: four one-dimensional representations and two two-dimensional representations. For the one-dimensional representations, the generators $\mathcal T$ and $\mathcal R$ effectively become two numbers $\sigma_{\mathcal T}$ and $ \sigma_{\mathcal R}$, so such a representation has the notation $\mathbf{1}_{\sigma_{\mathcal T} \sigma_{\mathcal R}}$. 
The two-dimensional irreducible representations are denoted by $\mathbf{2}_1$ and $\mathbf{2}_2$, which have the explicit representation matrices:
\begin{equation}
\mathbf{2}_1:\quad\mathcal{T}\mapsto
\begin{pmatrix}
\frac{1}{2}&\frac{3}{2}\\
-\frac{1}{2}&\frac{1}{2}
\end{pmatrix},\quad\mathcal{R}\mapsto
\begin{pmatrix}
1&0\\
0&-1
\end{pmatrix}
\end{equation}
\begin{equation}
\mathbf{2}_2:\quad\mathcal{T}\mapsto
\begin{pmatrix}
-\frac{1}{2}&\frac{3}{2}\\
-\frac{1}{2}&-\frac{1}{2}
\end{pmatrix},\quad\mathcal{R}\mapsto
\begin{pmatrix}
1&0\\
0&-1
\end{pmatrix}
\end{equation}


With standard techniques of group representation theory, the decomposition reads
\begin{equation}
\rm{span}\{\log(W_{290}),\ldots,\log(W_{313})\} =
4\,\mathbf{1}_{++}
\oplus
\,\mathbf{1}_{-+}
\oplus
\,\mathbf{1}_{+-}
\oplus
2\,\mathbf{1}_{--}
\oplus
3\,\mathbf{2}_1
\oplus
5\,\mathbf{2}_2 .
\end{equation}




\subsection{Symbology}
With the alphabet and the analytic canonical differential equation, we can compute the symbols of the pure basis for the DP and HB families, to higher weights. 

Recall that the symbol is defined as follows: for a pure function \(F^{(n)}\) of weight \(n\), the symbol is defined recursively by
\begin{equation}
dF^{(n)}=\sum_k F_k^{(n-1)}\,\mathrm d\log R_k
\qquad\Longrightarrow\qquad
\mathcal{S}\!\left(F^{(n)}\right)=\sum_k \mathcal{S}\!\left(F_k^{(n-1)}\right)\otimes R_k\,.
\end{equation}
Applied to the canonical differential equation system
\begin{equation}
d\vec{I}(\epsilon,\vec{x})
=
\epsilon\left(\sum_{i=1}^{N} \alpha_i\,\mathrm d\log W_i(\vec{x})\right)\vec{I}(\epsilon,\vec{x})\,,
\qquad
\vec{I}(\epsilon,\vec{x})=\epsilon^{-k}\sum_{m\ge 0}\epsilon^m \vec{I}^{(m)}(\vec{x})\,,
\end{equation}
this gives
\begin{equation}
\mathcal{S}\!\left(\vec{I}^{(m)}\right)
=\sum_{i_1,\ldots,i_m=1}^{N}
\alpha_{i_m}\cdots \alpha_{i_1}\,\vec{I}^{(0)}\,
S[W_{i_1},W_{i_2},\ldots,W_{i_m}]\,,\quad m>0
\end{equation}
with
\begin{equation}
    \mathcal{S}\!\left(\vec{I}^{(0)}\right) =\vec{I}^{(0)}\,,
\end{equation}
where \(S[W_{i_1},\ldots,W_{i_m}]\equiv W_{i_1}\otimes\cdots\otimes W_{i_m}\). In other words, once the constant matrices \(\alpha_i\)'s and the weight-zero values are known, the symbol of every pure master integral is generated algebraically.

Using the full differential equation in this work and the weight-zero values in~\cite{Henn:2025xrc}, we determine the symbols of the DP and HB families through weight six. In principle, with more RAM and computational time, these integrals' symbols can be evaluated at any weight.

We confirm the validity of the extended Steinmann relations~\cite{Caron-Huot:2019bsq} for all integrals in the basis of DP and HB families. These relations state that double discontinuities in partially overlapping channels vanish~\cite{Steinmann:1960a, Steinmann:1960b} and require that the letters corresponding to $s_{123}, s_{234}$ and $s_{345}$ never appear next to each other in the symbol. Physically, they reflect the incompatibility of the different three--particle cuts on all possible Riemann sheets. The differential equation matrices $\tilde{A}$ ensure that the Steinmann relations hold at any order in $\eps$ at any depth into the symbol \cite{Chicherin:2020umh}
\begin{equation}
    \alpha_{W_j} \cdot \alpha_{W_k}=0, \quad j \neq k, \quad j,k=7,8,9.
    \label{eq:adjacency}
\end{equation}
In practice, these adjacency conditions imply that any two letters $W_j$ and $W_k$ for which this identity holds will never appear as consecutive letters in the symbol. From the canonical differential equation systems obtained via the dihedral group action, we observe that $22433$ pairs of letters can never be adjacent in the symbol at any weight.

For the DP family's top sector pure integrals $I_i^\text{DP}$, $i=1,\ldots 5$,  nonzero symbols start to appear at weight five, four, five, six, and four, respectively. This is expected from the numerator's infrared structure and dual conformal symmetry.  

The double pentagon integral in $D=6-2\epsilon$, $I_4^\text{DP}$ starts from weight six. The weight-six part contains the following letters which are associated with $F_4$
\begin{equation}
W_{293},W_{296},W_{299},W_{302},W_{305},W_{308},W_{311}\,.
\end{equation}
Note that the weight-six part of $I_4^\text{DP}$ does {\it not} contain the letter $W_{290}=F_4$, because $W_{290}$ only appears in the diagonal entry $(4,4)$ of the differential equation. Since the weight-five part of $I_4^\text{DP}$ is zero, the weight-six part cannot contain $W_{290}$. The weight-$6$ symbol of $I_4^\text{DP}$ is available at
\begin{quotation}\url{https://bitbucket.org/yzhphy/2l6p_complete_function_space/src/main/version_2026/symbol/DP-I4w6.tar.xz}\ .
\end{quotation}

For the same reason, the dependence of all other integrals, $I_i^\text{DP}$, $i\not=4$, on $W_{290}\sim W_{313}$ can only come from the differential equation matrix elements $(i,4)$, $i\not=4$. Therefore, up to weight six, these integrals are free of the letters $W_{290}\sim W_{313}$. 

Furthermore, both the integrand of $I_4^\text{DP}$ and the constant $F_4$ factor have a $Z_2\times Z_2$ symmetry, i.e., left-right and up-down flips. In our notation, the generator \(\mathcal{R}\) in the dihedral group \(D_{6}\) corresponds to the left-right flip of the DP diagram, namely \(p_{i}\rightarrow p_{7-i}\), while the element \(\mathcal{R}\mathcal{T}^3\) is the up-down flip \(p_{1}\leftrightarrow p_{3}\) and \(p_{4}\leftrightarrow p_{6}\). What we have explicitly checked is that 
the weight-six symbol of \(I_{4}^\text{DP}\) respects this \(Z_{2}\times Z_{2}\) symmetry.

We have an interesting observation: the weight-six components of two-loop six-point integrals, with the dihedral group action, contain {\it all} the $245$ letters defined in \cite{Abreu:2024fei,Henn:2025xrc}, as well as $21$  new letters $W_{293} \sim W_{313}$ in \eqref{eq:NewF4odd}.  As a comparison, up to the weight-four, two-loop six-point integrals with the dihedral group action, contain only a subset ($167$ letters) out of the $245$ letters~\cite{Abreu:2024fei,Henn:2025xrc}.


\section{Numerical Evaluation of the Differential Equation}
\label{section:numerical evaluation}
With the complete analytic differential equations, we can use them to numerically evaluate the master integrals at different kinematic points. As a proof of concept for fast numerical evaluation, we use the method of Chebyshev pseudospectral transport to numerically evaluate DP integrals up to the weight six in the physical scattering region.

\subsection{Kinematic regions and boundary values}
After determining the complete set of genuine letters, the canonical system of differential equations is fully specified. A natural reference sheet for six-point kinematics is the Euclidean region~\cite{Henn:2024ngj}. In the all-outgoing convention, and using the parity-even variables
\begin{equation}
\vec{s}=\{s_{12},s_{23},s_{34},s_{45},s_{56},s_{61},s_{123},s_{234},s_{345}\}\,,
\end{equation}
we define the Euclidean region by
\begin{equation}
s_{12},\,s_{23},\,s_{34},\,s_{45},\,s_{56},\,s_{61},\,s_{123},\,s_{234},\,s_{345}<0\,.
\end{equation}
For four-dimensional external kinematics, this should be understood as the connected component of the real Gram surface,
\begin{equation}
G(1,2,3,4,5)\equiv \det\!\bigl(2\,p_i\!\cdot p_j\bigr)_{1\le i,j\le 5}=0\,,
\end{equation}
that contains the symmetric point
\begin{equation}
\vec{s}_{E}=\{-1,\ldots,-1\}\,.
\end{equation}
In this domain, dimensionally regulated Feynman integrals are real-valued. The differential-equation matrices may nevertheless exhibit spurious singular hypersurfaces at zeros of alphabet letters, which are not branch points of the integrals themselves but must be avoided when choosing numerical contours~\cite{Henn:2024ngj}. The boundary values and the evaluation of the two-loop six-point integrals' differential equation, up to the weight four were discussed in~\cite{Abreu:2024fei, Henn:2024ngj}. Thus, in this section we take the physical region as an example of how to evaluate the two-loop six-point planar massless integrals up to weight six, since the physical region is directly associated with the particle scattering processes.

In order to specify a physical scattering region, consider the scattering channel $12\to3456$. In the all-outgoing convention and with Minkowski signature, the on-shell kinematics require
\begin{equation}
\label{eq:sign-2to4}
  s_{12},\, s_{34},\, s_{35},\, s_{36},\, s_{45},\, s_{46},\, s_{56} > 0\,,
  \qquad
  s_{13},\, s_{14},\, s_{15},\, s_{16},\, s_{23},\, s_{24},\, s_{25},\, s_{26} < 0\,.
\end{equation}
The physical scattering region also requires: 
\begin{equation}
G(1,2,3,4,5)=0,
\qquad
 G(i,j,k,l)<0, \ \text{distinct } i,j,k,l,
\qquad
G(i,j,k)>0, \ \text{distinct } i,j,k,
\label{eq:gram_sign-2to4}
\end{equation}
Hence the physical $12\to3456$ region is defined by the constraints~\eqref{eq:sign-2to4} and \eqref{eq:gram_sign-2to4} which contains the sample points $\vec{x}_0$ and $\vec{x}_1$.


In this work, we choose two kinematic points in this physical region, with the help of a private version of momentum twistor parametrization\footnote{We thank Dmitry Chicherin for sharing his private parametrization. In this parametrization, when the eight independent variables are real numbers, with a high probability, the kinematic point is in the physical scattering region. The constraint $G(1,2,3,4,5)=0$ holds automatically and all pseudo-scalars are rationalized.}
\begin{equation}
\begin{aligned}
\vec{x}_0 &= \left\{\frac{325}{16},-\frac{181}{16},\frac{65}{8},\frac{9}{4},\frac{19}{8},-\frac{19}{16},7,-\frac{109}{16},\frac{115}{8}\right\}\,,\\
\vec{x}_1 &= \left\{\frac{41202613}{1969920},-\frac{433189109}{37696000},\frac{5365589}{634880},\frac{403880939}{152668800},\frac{57623740097}{24427008000},\right.\\
&\qquad \left.-\frac{1115291}{1036800},\frac{10965881791}{1526688000},-\frac{2160073789}{301568000},\frac{30049703}{1969920}\right\}\,.
\end{aligned}
\end{equation}

High-precision boundary values at $\vec{x}_0$ are obtained using \texttt{AMFlow}, with about 200-digit numerical accuracy. The \texttt{AMFlow} run is accelerated by choosing a scheme in which all external legs are four-dimensional, as suggested in Ref.~\cite{Bargiela:2025nqc}. However, this step is still computationally expensive and requires approximately $72\,\mathrm{hrs}$, but it is performed only once. The essential point is that, once a single seed point in the  physical region is known, the canonical system should transport all master integrals rapidly to nearby points. 

In this work, we demonstrate the correctness of the analytic canonical differential equation using a new numerical solver. The new solver, powered by Chebyshev pseudospectral transport, is extremely fast and capable of reaching a high precision. We also remark that the singularity analysis of the $2\to 4$ scattering region is complicated and is left for future work. In this work, the points $\vec{x}_0$ and $\vec{x}_1$ are in close proximity to one another, and the path can be chosen to be free of singularities.

\subsection{Chebyshev pseudospectral transport of the canonical system}
\label{subsec:rk-transport}

Our new implementation replaces the traditional contour integration by a global Chebyshev pseudospectral solve~\cite{boyd2001chebyshev,reddy2005accuracy}. After choosing a contour $\gamma:[0,1]\to\mathbb{C}^9$ with $\gamma(0)=\vec{x}_0$ and $\gamma(1)=\vec{x}_1$, and a coordinate $t$ for $[0,1]$, the multivariate canonical system reduces to the one-dimensional equation
\begin{equation}
\frac{\mathrm d}{\mathrm dt}\vec{I}(\epsilon,t)
=
\epsilon\,A(t)\,\vec{I}(\epsilon,t)\,,
\qquad
A(t)=\sum_{W_i \in \mathbb A} \alpha_{W_i}\,\frac{\mathrm d}{\mathrm dt}\log W_i(\gamma(t))\,,
\end{equation}
where the matrices $\alpha_{W_i}$ are rational constants and all kinematic dependence is encoded in the letters. Expanding the master integrals in a Laurent series around $\epsilon=0$
\begin{equation}
\vec{I}(\epsilon,t)=\epsilon^{-4} \sum_{r=0}^{6}\epsilon^r\,\vec{I}^{(r)}(t)\,,
\end{equation}
one obtains the triangular hierarchy
\begin{equation}
\frac{\mathrm d}{\mathrm dt}\vec{I}^{(r)}(t)=A(t)\,\vec{I}^{(r-1)}(t)\,,
\qquad r=1,\ldots,6\,,
\end{equation}
with $\vec{I}^{(0)}(t)$ constant along the contour.

It is convenient to collect the seven transported Laurent coefficients into the block vector
\begin{equation}
\mathbf{Y}(t)=
\begin{pmatrix}
\vec{I}^{(0)}(t)\\
\vec{I}^{(1)}(t)\\
\vdots\\
\vec{I}^{(6)}(t)
\end{pmatrix},
\end{equation}
so that the full transport problem becomes a single linear system
\begin{equation}
\mathbf{Y}'(t)=\mathcal{A}(t)\,\mathbf{Y}(t)\,,
\qquad
\mathcal{A}(t)=
\begin{pmatrix}
0 & 0 & \cdots & 0 \\
A(t) & 0 & \cdots & 0 \\
0 & A(t) & \ddots & \vdots \\
\vdots & \ddots & \ddots & 0 \\
0 & \cdots & A(t) & 0
\end{pmatrix}.
\end{equation}
This sparse block-subdiagonal operator is exactly the object constructed by the Mathematica routine \texttt{RecursiveOperatorMatrix} in the file \texttt{oneline\_CL.wl}, which is available at,
\begin{quotation}
\url{https://bitbucket.org/yzhphy/2l6p_complete_function_space/src/main/version_2026/evaluation/}
\end{quotation}

We then approximate $\mathbf{Y}(t)$ on the full interval $[0,1]$ by its degree-$m$ Chebyshev interpolant through the Chebyshev--Lobatto nodes
\begin{equation}
t_j=\frac{1}{2}\!\left(1-\cos\frac{j\pi}{m}\right),\qquad j=0,\ldots,m\,.
\end{equation}
If $\mathbf{Y}_j$ denotes the approximation to $\mathbf{Y}(t_j)$, the derivative of the interpolant at the nodes is represented by the Chebyshev differentiation matrix $D$
\begin{equation}
\mathbf{Y}'(t_j)\approx \sum_{\ell=0}^{m} D_{j\ell}\,\mathbf{Y}_{\ell}\,.
\end{equation}
The initial condition is imposed exactly at the left endpoint
\begin{equation}
\mathbf{Y}_0=\mathbf{Y}(0)\,,
\end{equation}
while at the remaining nodes we enforce the differential equation
\begin{equation}
\sum_{\ell=0}^{m} D_{j\ell}\,\mathbf{Y}_{\ell}-\mathcal{A}(t_j)\,\mathbf{Y}_{j}=0\,,
\qquad j=1,\ldots,m\,.
\end{equation}
This produces a single sparse linear system of dimension $(m+1)N_{\rm state}$, where $N_{\rm state}$ is the dimension of $\mathbf{Y}(t)$. Solving that system yields the transported solution simultaneously at all collocation nodes and, in particular, at the endpoint $t=1$. In the implementation, \texttt{ChebyshevLobattoData} constructs the mapped Lobatto grid and the differentiation matrix, while \texttt{SpectralPropagate} assembles the sparse collocation equations and solves them with \texttt{LinearSolve}.

The crucial advantage of this approach is spectral convergence. As long as the contour stays away from zeros of the letters, the one-dimensional connection $A(t)$ and the transported solution are analytic in a neighborhood of the interval, and the collocation error decreases exponentially with the number of nodes. In contrast to local step-by-step integration, the pseudospectral method exploits the global smoothness of the canonical system along the entire contour. This is especially effective here because the nontrivial kinematic dependence enters only through the scalar functions $\partial_t\log W_a$, whereas the matrices $\alpha_{W_i}$ are constant and the coupling in Laurent-coefficient space is strictly lower triangular. The resulting numerical problem is therefore reduced to sparse linear algebra at a small number of collocation points.
\begin{table}[htbp]
\centering
\small
\setlength{\tabcolsep}{5pt}
\renewcommand{\arraystretch}{1.15}
\begin{tabular}{@{}cc>{\raggedleft\arraybackslash}p{4.9cm}c@{}}
\toprule
Integral & Weight & \texttt{AMFlow} value at $\vec{x}_1$
& $|\Delta|$  \\
\midrule
\multirow{2}{*}{$I_1$}
& 5 & \makecell[r]{$-2.46925799967801021$\\$-1.96660747417995791\,i$}
&  $7.7074892561505088\times 10^{-35}$ \\
& 6 & \makecell[r]{$\phantom{-}11.3971967377661105$\\$-28.5325963721922041\,i$}
&   $6.2828281332144624\times 10^{-35}$\\
\midrule
\multirow{2}{*}{$I_2$}
& 5 & \makecell[r]{$-1.12562474681489852$\\$+3.14851951381445006\,i$}
&  $1.6292682378716045\times 10^{-35}$ \\
& 6 & \makecell[r]{$-18.9231030042537387$\\$+6.30164562806054788\,i$}
&   $1.4233575451604684\times 10^{-34}$\\
\midrule
\multirow{2}{*}{$I_3$}
& 5 & \makecell[r]{$0$\\$-1.60391051639677169\,i$}
&   $1.7246963953181861\times 10^{-38}$\\
& 6 & \makecell[r]{$\phantom{-}8.88814980874023443$\\$-5.28697363215292301\,i$}
&   $1.8115956417466913\times 10^{-36}$\\
\midrule
\multirow{2}{*}{$I_4$}
& 5 & $0$
&   $6.4964814821739773\times 10^{-112}$\\
& 6 & \makecell[r]{$\phantom{-}0.264970209178792364$\\$-0.226755406317898571\,i$}
&   $2.1058509863411872\times 10^{-36}$\\
\midrule
\multirow{2}{*}{$I_5$}
& 5 & \makecell[r]{$-7.24249371419095952$\\$-12.9906469797979563\,i$}
&   $2.1402651151685830\times 10^{-35}$\\
& 6 & \makecell[r]{$\phantom{-}10.3500603199039786$\\$-44.9314387523581872\,i$}
&   $2.3520461307392542\times 10^{-34}$\\
\midrule
\multicolumn{4}{r}{\textbf{max. absolute discrepancy shown $2.3520461307392542\times 10^{-34}$}}  \\
\bottomrule
\end{tabular}
\caption{Benchmark comparison between the (24 nodes) Chebyshev pseudospectral transport method and \texttt{AMFlow} for the weight-five and weight-six coefficients of the five top-sector master integrals at $\vec{x}_1$. We define $\Delta\equiv I^{\rm method}-I^{\rm AMFlow}$. }
\label{tab:rk-amflow-top}
\end{table}

\begin{table}[htbp]
\centering
\small
\renewcommand{\arraystretch}{1.15}
\begin{tabular}{@{}cccc@{}}
\toprule
\#Chebyshev nodes & Observed max-abs agreement at $\vec{x}_1$ & Runtime & Memory \\
\midrule
12&$\sim 7.58\times10^{-16}$ & $18.09\,\mathrm{s}$ & $2.9\,\mathrm{GB}$ \\
24&$\sim 1.36\times10^{-31}$ & $144.76\,\mathrm{s}$ & $26.1\,\mathrm{GB}$\\
48&$\sim 1.96\times10^{-62}$ & $228.42\,\mathrm{s}$ & $42.2\,\mathrm{GB}$\\
96&$< 1\times 10^{-100}$ & $1895.34\,\mathrm{s}$ & $\sim 200\,\mathrm{GB}$\\
192&$<1 \times 10^{-100}$ & $13739\,\mathrm{s}$ & $\sim 1400\,\mathrm{GB}$\\
\bottomrule
\end{tabular}
\caption{Performance of the Chebyshev pseudospectral transport from $\vec{x}_0$ to $\vec{x}_1$. The quoted accuracy refers to the maximum absolute discrepancy with respect to the independent \texttt{AMFlow} reference values (100-digit precision) at $\vec{x}_1$. The cases with 12--48 nodes were evaluated using a single core of an Intel 13th Gen Core i9-13950HX with 128 \rm{GB} memory, while 96 and 192 nodes evaluation were performed on single core of an Intel Xeon Gold 6330 with 1.97 \rm{TB} memory. }
\label{tab:rk-amflow-top2}
\end{table}

In practice, our Mathematica implementation \texttt{oneline\_CL.wl} follows exactly this strategy. For the benchmark considered here, the transport from $\vec{x}_0$ to $\vec{x}_1$ reaches the accuracy levels summarized in Table~\ref{tab:rk-amflow-top}. This comparison should therefore be viewed as more than a numerical cross-check: once a single high-precision seed point on the correct physical sheet has been fixed, the canonical differential equation itself becomes a fast, systematically improvable, and very high-precision numerical representation of the master integrals in the physical region, see Table~\ref{tab:rk-amflow-top2}.

For the present physical-region benchmark, alternative approaches based on explicit one-fold or two-fold integral representations~\cite{Henn:2024ngj,Liu:2024ont} would require a more involved setup. By contrast, the Chebyshev pseudospectral transport makes direct use of the reconstructed 269-letter canonical system and turns the evaluation problem into a sparse linear-algebra problem on a small set of collocation nodes. In our tests, a direct use of the public code \texttt{DiffExp}~\cite{Hidding:2020ytt} did not lead to comparably stable high-precision results along the chosen contour, which further underscores the advantage of the global spectral treatment for this system.

\section{Summary and Outlook}
\label{section:summary}

In this paper, we extended the analytic study of planar massless two-loop six-point Feynman integrals beyond the previously known weight-four level and into higher orders in the dimensional regulator~$\epsilon$. 
The central new ingredient is the genuinely nine-propagator double-pentagon sector. While this sector does not generate independent weight-four functions in the four-dimensional analyses of Refs.~\cite{Abreu:2024fei,Henn:2025xrc}, it plays an essential role beyond the finite part. 
By analytically reconstructing the previously missing off-diagonal part of the canonical differential equation, we obtained the canonical system needed for the higher-order analysis of planar two-loop six-point integrals.

A main outcome of this work is the corresponding extension of the symbol alphabet. 
Starting from the $245$ letters relevant up to weight four in Refs.~\cite{Abreu:2024fei,Henn:2025xrc}, and combining them with the $18$ letters associated with the diagonal block of the double-pentagon top sector, we identified six genuinely new letters associated with the $D=6-2\epsilon$ double-pentagon master integral. 
Altogether, this yields $269$ independent letters for planar massless two-loop six-point integrals. 
Using this system, together with the weight-zero data of Ref.~\cite{Henn:2025xrc}, we determined the symbols of the double-pentagon and hexagon-box families through weight six, i.e.\ through $\mathcal{O}(\epsilon^2)$, and demonstrated high-precision numerical evaluation in a representative physical scattering region.

From the amplitude point of view, this higher-$\epsilon$ information is  the type of input expected to be relevant for future studies of the infrared structure and function space of planar three-loop six-point scattering.  In this sense, our results complement the finite-part constructions of Refs.~\cite{Abreu:2024fei,Henn:2025xrc}  by supplying data that are invisible at weight four but become useful in higher-order applications. 
They should also provide concrete higher-weight input for studies of symbology and cluster-algebraic structures.

In addition, we showed that once the canonical system is known, it can be turned directly into an efficient numerical representation in physical kinematics. 
The Chebyshev pseudospectral transport introduced here provides a robust alternative to more traditional numeric differential equation methods, and appears well suited to large multi-scale systems. The method offers an efficient approach to numerical integration: using only a single CPU core for less than one minute, it obtains results for weight six at standard double-precision floating-point accuracy, and by simply increasing the number of Chebyshev--Lobatto nodes, it can deliver highly accurate numerical evaluations up to $10^{-60}$ precision within five minutes. This may provide a practical tool for future higher-loop calculations in nontrivial kinematic regions.

A natural next step is to combine the present canonical differential system with a systematic analysis of the full $2\to4$ physical region and to lift the weight-six symbol data to function-level representations. 
It will also be interesting to explore further how the extended alphabet and higher-weight data obtained here can constrain bootstrap approaches to six-point three-loop observables and sharpen the study of cluster-algebraic structures. 

\section*{Acknowledgments}
We are deeply grateful to Johannes Henn for his valuable contributions in the early stages of this work and for his insightful suggestions throughout the project. We warmly thank Dmitry Chicherin for generously providing us with his private version of the momentum twistor parametrization, which proved essential for our computations in the physical region. We also thank James Drummond and Ömer Gürdoğan for their valuable comments on the algebraic structure of the new letters. Finally, we extend our sincere thanks to Samuel Abreu, Rigers Aliaj, Piotr Bargiela, David Kosower, Pier Monni, Georgios Papathanasiou, Marcus Spradlin, Anastasia Volovich, Stefan Weinzierl and  Qinglin Yang for the many helpful discussions and insightful comments.

The research of Yuanche Liu is supported by NSFC through Grant No. 124B1014. Yingxuan Xu is supported by the Deutsche
Forschungsgemeinschaft (DFG, German Research Foundation) under grant 396021762 - TRR 257. Yang Zhang is supported by NSFC through Grant No. 12575078 and 12247103. Two of the authors (AM and YZ) would like to thank the Erwin Schrödinger International Institute for Mathematics and Physics (ESI), University of Vienna (Austria), for the opportunity to participate in the Thematic Programme ``Amplitudes and Algebraic Geometry" in 2026 where a significant part of this work has been accomplished and for the support given.

\appendix
\section{Two--loop Six--point Alphabet, up to the weight four}
\label{app:alphabet}

In this Appendix, we recall the two--loop six--point alphabet from~\cite{Henn:2025xrc}. The alphabet consists of 48 letters linear in Mandelstam variables $s_{ij}$ and 51 quadratic letters:
\begin{align}
    W_{1}&=s_{12}, & W_{i+1}&=\mathcal{T}^{i} (W_{1}), & i&=1,\ldots,5, \\
    W_{7}&=s_{123}, & W_{i+7}&=\mathcal{T}^{i} (W_{7}), & i&=1,2, \\
    W_{10}&=-s_{12}-s_{23}, & W_{i+10}&=\mathcal{T}^{i} (W_{10}), &i&=1,\ldots,5, \\
    W_{16}&=s_{12}-s_{123}, & W_{i+16}&=\mathcal{T}^{i} (W_{16}), &i&=1,\ldots,5, \\
    W_{22}&=s_{12}-s_{345}, & W_{i+22}&=\mathcal{T}^{i} (W_{22}), &i&=1,\ldots,5, \\
    W_{28}&=-s_{12}-s_{23}+s_{123}, & W_{i+28}&=\mathcal{T}^{i} (W_{28}), &i&=1,\ldots,5, \\
    W_{34}&=s_{12}-s_{34}-s_{123}, & W_{i+34}&=\mathcal{T}^{i} (W_{34}), &i&=1,\ldots,5, \\
    W_{40}&=s_{12}-s_{56}-s_{345}, & W_{i+40}&=\mathcal{T}^{i} (W_{40}), & i&=1,\ldots,5, \\
    W_{46}&=s_{12}+s_{45}-s_{123}-s_{345}, & W_{i+46}&=\mathcal{T}^{i} (W_{46}), &i&=1,2, \\
    W_{49}&=-s_{12}s_{45}+s_{123}s_{345}, 
    &W_{i+49}&=\mathcal{T}^{i} (W_{49}),  &i&=1,2, \\
    W_{52}&=s_{12} s_{56}-s_{12} s_{123}+s_{34} s_{123} , &W_{i+52}&=\mathcal{T}^{i}
    (W_{52}), &i&=1,\ldots,5, \\
    W_{58}&=-s_{12}s_{45}-s_{23}s_{345}+s_{123}s_{345}, 
    &W_{i+58}&=\mathcal{T}^{i} (W_{58}), &i&=1,\ldots,5, \\
    W_{64}&=-s_{12}s_{45}-s_{34}s_{123}+s_{123}s_{345},  &W_{i+64}&=\mathcal{T}^{i} (W_{64}), &i&=1,\ldots,5, \\
    W_{70}&=s_{12} s_{56}-s_{123} s_{56}+s_{34} s_{123},  &W_{i+70}&=\mathcal{T}^{i} (W_{70}),  &i&=1,\ldots,5, \\
    W_{76}&=s_{12}(s_{34}+s_{45})-(s_{34}-s_{56}+s_{123})s_{345}, &W_{i+76}&=\mathcal{T}^{i}
    (W_{76}), &i&=1,\ldots,5, \\
    W_{82}&=(s_{12}+s_{23})s_{45}+s_{123}(s_{61}-s_{23}-s_{345}), &W_{i+82}&=\mathcal{T}^{i} (W_{82}),&i&=1,\ldots,5, \\
    W_{88}&=s_{12}(-s_{34}+s_{345})+(s_{34}-s_{56}-s_{345})s_{345}
, &W_{i+88}&=\mathcal{T}^{i} (W_{88}), &i&=1,\ldots,5, \\
    W_{94}&=(s_{34}-s_{12}+s_{123})(s_{12}-s_{34}+s_{234})-s_{23}s_{56}, 
    &W_{i+94}&=\mathcal{T}^{i} (W_{94}),  &i&=1,\ldots,5. 
\end{align}
Additionally, there are 18 cubic letters: 
\begin{align}
    W_{100}&=s_{23}s_{56}(-s_{34}+s_{345})-(s_{61}-s_{234})(s_{12}s_{45}+s_{34}s_{123}-s_{123}s_{345}), \notag \\
    W_{i+100}&=\mathcal{T}^{i} (W_{100}), \quad i=1,\ldots,5\\
    W_{106}&=-s_{123}\left(s_{34}-s_{56}\right)  \left(s_{56}-s_{234}\right)-s_{12} s_{56} \left(s_{23}+s_{56}-s_{234}\right), \notag \\
    W_{i+106}&=\mathcal{T}^{i} (W_{106}), \quad i=1,\ldots,5 \\
    W_{112}&=\left(s_{234}-s_{61}\right) s_{23}^2-\left(s_{56} s_{61}-s_{123} s_{61}+\left(s_{45}+s_{123}\right) s_{234}\right) s_{23}+s_{45} s_{123} s_{234}, \notag \\
    W_{i+112}&=\mathcal{T}^{i} (W_{112}), \quad i=1,\ldots,5.
\end{align}
The following five letters are square roots in both Mandelstam variables $s_{ij}$ and in momentum twistor variables $x_i$:
\begin{align}
    W_{118}&=r_1=\sqrt{\lambda(s_{12},s_{34},s_{56})}, \\
    W_{119}&=r_2=\sqrt{\lambda(s_{23},s_{45},s_{61})}, \\
    W_{120}&=r_3=\sqrt{\lambda(s_{12},s_{36},s_{45})}, \notag \\
     W_{i+120}&=r_{i+3}=\mathcal{T}^{i} (W_{120}), \quad i=1,2,
\end{align}
where $\lambda$ denotes the K\"all\'en function defined in eq.~\eqref{eq:Kallen}.

Next, we have 34 letters involving pseudo--scalars. All of these letters are square roots in Mandelstam variables, but become perfect squares in momentum twistor variables.
\begin{align}
    W_{123}&=\epsilon_{1234},\qquad  W_{i+123}=\mathcal{T}^{i} (W_{123}), \qquad i=1,\ldots,5, \\ 
    W_{129}&=\epsilon_{1235}, \qquad  W_{i+129}=\mathcal{T}^{i} (W_{129}),  \qquad i=1,\ldots,5, \\ 
    W_{135}&=\epsilon_{1245},\qquad   W_{i+135}=\mathcal{T}^{i} (W_{135}), \qquad i=1,2, \\ 
    W_{138}&=\Delta_6\\
    W_{139}&=s_{12}\epsilon_ {1456} + s_{123} \epsilon_ {1256},\qquad  W_{i+139}=\mathcal{T}^{i} (W_{139}),\qquad  i=1,\ldots,5, \\ 
    W_{145}&=s_{56} \epsilon_ {1234} - 
 s_{34} \epsilon_ {1245} - (s_{34} - s_{234})\epsilon_ {1345} - (s_{56} - 
     s_{234})\epsilon_ {2345}, \notag \\
    W_{i+145}&=\mathcal{T}^{i} (W_{145}),\qquad  i=1,\ldots,5, \\ 
    W_{151}&= s_{234} \epsilon_{6123}-s_{61} \epsilon_{1234},\qquad  W_{i+151}=\mathcal{T}^{i} (W_{151}),\qquad  i=1,\ldots,5.
\end{align}
All of the letters listed so far, are so-called even letters since they are invariant under sign flipping of square roots $\sqrt{r}\to - \sqrt{r}$ or parity transformation of pseudo-scalars $\eps_{ijkl}\to - \eps_{ijkl}$.

The remaining letters are so-called odd letters since they change sign under parity transformations $\log(W_i)\to -\log(W_i)$. First, we have 25 letters which are odd with respect to square roots $r_i$:
\begin{align}
   W_{157}&=\dfrac{s_{12}+s_{34}-s_{56}-r_1}{s_{12}+s_{34}-s_{56}+r_1}, &  W_{158}&=\mathcal{T} (W_{157}), \\ 
   W_{159}&=\dfrac{-s_{12}+s_{34}+s_{56}-r_1}{-s_{12}+s_{34}+s_{56}+r_1}, &  W_{160}&=\mathcal{T}(W_{159}), \\ 
   W_{161}&=\dfrac{s_{12}-s_{34}+s_{56}-2s_{123}-r_1}{s_{12}-s_{34}+s_{56}-2s_{123}+r_1}, & W_{i+161}&=\mathcal{T}^{i} (W_{161}), & i&=1,\ldots,5, \\ 
   W_{167}&=\dfrac{s_{123}+s_{345}-r_3}{s_{123}+s_{345}+r_3}, & W_{i+167}&=\mathcal{T}^{i} (W_{167}), & i&=1,2, \\ 
   W_{170}&=\dfrac{s_{123}-s_{345}-r_3}{s_{123}-s_{345}+r_3}, & W_{i+170}&=\mathcal{T}^{i} (W_{170}), & i&=1,2, \\ 
   W_{173}&=\dfrac{s_{123}+s_{345}-2s_{12}-r_3}{s_{123}+s_{345}-2s_{12}+r_3}, & W_{i+173}&=\mathcal{T}^{i} (W_{173}), & i&=1,2, \\ 
   W_{176}&=\dfrac{s_{123}-s_{345}+2s_{34}-2s_{56}-r_3}{s_{123}-s_{345}+2s_{34}-2s_{56}+r_3}, & W_{i+176}&=\mathcal{T}^{i} (W_{176}), & i&=1,2\\
   W_{179}&=\dfrac{s_{123}-s_{345}-2s_{23}+2s_{61}-r_3}{s_{123}-s_{345}-2s_{23}+2s_{61}+r_3}, & W_{i+179}&=\mathcal{T}^{i} (W_{179}), & i&=1,2.
\end{align}
Second, we have 93 letters that are odd with respect to parity transformation of pseudo-scalars and $\Delta_6$:
\begin{align}
    W_{182}&=\dfrac{s_{23} \left(s_{56}-s_{34}\right)+s_{34} s_{123}+s_{12} \left(s_{23}-s_{234}\right)-s_{123} s_{234}-\eps_{1234}}{s_{23} \left(s_{56}-s_{34}\right)+s_{34} s_{123}+s_{12} \left(s_{23}-s_{234}\right)-s_{123} s_{234}+\eps_{1234}}, \notag \\
    W_{i+182}&=\mathcal{T}^{i} (W_{182}), \qquad i=1,\ldots,5, \\
    W_{188}&=\dfrac{s_{12}(s_{23}-s_{234})-s_{23}(s_{34}+s_{56})+s_{123}(s_{34}+s_{234})-\eps_{1234}}{s_{12}(s_{23}-s_{234})-s_{23}(s_{34}+s_{56})+s_{123}(s_{34}+s_{234})+\eps_{1234}}, \notag \\
    W_{i+188}&=\mathcal{T}^{i} (W_{188}), \qquad i=1,\ldots,5,\\
    W_{194}&=\dfrac{s_{12}(-s_{23}+s_{234})+s_{23}(s_{34}+s_{56})+s_{123}(s_{34}-s_{234})-\eps_{1234}}{s_{12}(-s_{23}+s_{234})+s_{23}(s_{34}+s_{56})+s_{123}(s_{34}-s_{234})+\eps_{1234}}, \notag \\
    W_{i+194}&=\mathcal{T}^{i} (W_{194}), \qquad i=1,\ldots,5, \\
    W_{200}&=\dfrac{s_{12}(s_{23}+s_{234})+s_{23}(-s_{34}+s_{56})+s_{123}(s_{34}-s_{234})-\eps_{1234}}{s_{12}(s_{23}+s_{234})+s_{23}(-s_{34}+s_{56})+s_{123}(s_{34}-s_{234})+\eps_{1234}}, \notag \\
    W_{i+200}&=\mathcal{T}^{i} (W_{200}), \qquad i=1,\ldots,5,\\
    W_{206}&=\dfrac{s_{12}(s_{23}+s_{234})-s_{23}(s_{34}+s_{56}-2s_{234})+s_{123}(s_{34}-s_{234})-\eps_{1234}}{s_{12}(s_{23}+s_{234})-s_{23}(s_{34}+s_{56}-2s_{234})+s_{123}(s_{34}-s_{234})+\eps_{1234}}, \notag \\
    W_{i+206}&=\mathcal{T}^{i} (W_{206}), \qquad i=1,\ldots,5, \\
    W_{212}&=\dfrac{s_{12}(s_{23}-s_{234})+s_{23}(-s_{34}+s_{56}-2s_{123})+s_{123}(-s_{34}+s_{234})-\eps_{1234}}{s_{12}(s_{23}-s_{234})+s_{23}(-s_{34}+s_{56}-2s_{123})+s_{123}(-s_{34}+s_{234})+\eps_{1234}}, \notag \\
    W_{i+212}&=\mathcal{T}^{i} (W_{212}), \qquad i=1,\ldots,5, \\
    W_{218}&=\dfrac{s_{12}(s_{23}+2s_{56}-2s_{123}-s_{234})+s_{23}(-s_{34}+s_{56})+s_{123}(s_{34}-s_{234})-\eps_{1234}}{s_{12}(s_{23}+2s_{56}-2s_{123}-s_{234})+s_{23}(-s_{34}+s_{56})+s_{123}(s_{34}-s_{234})+\eps_{1234}}, \notag \\
    W_{i+218}&=\mathcal{T}^{i} (W_{218}), \qquad i=1,\ldots,5, \\
    W_{224}&=\dfrac{2s_{12}^2+s_{12}(s_{23}-2s_{34}-2s_{123}+s_{234})+s_{23}(-s_{34}+s_{56})+s_{123}(s_{34}-s_{234})-\eps_{1234}}{2s_{12}^2+s_{12}(s_{23}-2s_{34}-2s_{123}+s_{234})+s_{23}(-s_{34}+s_{56})+s_{123}(s_{34}-s_{234})+\eps_{1234}}, \notag \\
    W_{i+224}&=\mathcal{T}^{i} (W_{224}), \qquad i=1,\ldots,5,\\
    W_{230}&=-\dfrac{s_{12} \left(s_{23}+2 s_{56}-s_{234}\right)+s_{23} \left(s_{56}-s_{34}\right)+s_{123} \left(s_{34}-2 s_{56}+s_{234}\right)-\eps_{1234}}{s_{12} \left(s_{23}+2 s_{56}-s_{234}\right)+s_{23} \left(s_{56}-s_{34}\right)+s_{123} \left(s_{34}-2 s_{56}+s_{234}\right)+\eps_{1234}}, \notag \\
    W_{i+230}&=\mathcal{T}^{i} (W_{230}), \qquad i=1,\ldots,5, \\
    W_{236}&=\dfrac{s_{12} \left(s_{23}-s_{234}\right)-s_{23} \left(s_{34}+s_{56}\right)-\left(2 s_{56}-s_{123}\right) \left(s_{34}-s_{234}\right)-\eps_{1234}}{s_{12} \left(s_{23}-s_{234}\right)-s_{23} \left(s_{34}+s_{56}\right)-\left(2 s_{56}-s_{123}\right) \left(s_{34}-s_{234}\right)+\eps_{1234}}, \notag \\
    W_{i+236}&=\mathcal{T}^{i} (W_{236}), \qquad i=1,\ldots,5, \\
    W_{242}&=\dfrac{-s_{12}(s_{45}+s_{61}-s_{234})+s_{23}(s_{34}+s_{56}-s_{345})+s_{123}(-s_{34}+s_{61}-s_{234}+s_{345})-\eps_{1235}}{-s_{12}(s_{45}+s_{61}-s_{234})+s_{23}(s_{34}+s_{56}-s_{345})+s_{123}(-s_{34}+s_{61}-s_{234}+s_{345})+\eps_{1235}}, \notag \\
    W_{i+242}&=\mathcal{T}^{i} (W_{242}), \qquad i=1,\ldots,5,
\end{align}
\begin{align}
    W_{248}&=\dfrac{P_1-\eps_{1245}}{P_1+\eps_{1245}}, 
    &W_{i+248}&=\mathcal{T}^{i} (W_{248}), &i&=1,2, \\
    W_{251}&=\dfrac{P_2-\eps_{1245}}{P_2+\eps_{1245}},  
    &W_{i+251}&=\mathcal{T}^{i} (W_{251}), &i&=1,2, \\
    W_{254}&=\dfrac{P_3-\eps_{1245}}{P_3+\eps_{1245}}, 
    &W_{i+254}&=\mathcal{T}^{i} (W_{254}), &i&=1,2, \\
    W_{257}&=\dfrac{P_4-\eps_{1245}}{P_4+\eps_{1245}}, 
    &W_{i+257}&=\mathcal{T}^{i} (W_{257}), &i&=1,2,
\end{align}
where $P_1$, $P_2$, $P_3$ and $P_4$ are polynomials defined as
\begin{align}
    P_{1}&=s_{12}s_{45}+s_{23}(-s_{34}+s_{56}+s_{345})+s_{61}(s_{34}-s_{56})+s_{123}(s_{34}+s_{61}-s_{234}) \notag \\
    &\qquad +s_{345}(s_{56}-s_{123}-s_{234}),\\
    P_{2}&=s_{12}s_{45} - s_{34}s_{61} +s_{56}s_{61} - s_{34}s_{123}+s_{123}(-s_{61}+ s_{234} + s_{345}) \notag \\
   &\qquad + s_{23} (s_{34} -s_{56} - s_{345})-s_{56}s_{345} + s_{234}s_{345},\\
    P_{3}&=s_{12}s_{45}+s_{123}(-s_{34}+s_{234})+2s_{23}^2+s_{23}\left(s_{34}+s_{56}-2(s_{61}+s_{123}+s_{234})+s_{345}\right) \notag\\
    &\qquad +s_{61}(-s_{34}-s_{56}+s_{123}+2s_{234})+s_{345}(s_{56}-s_{123}-s_{234}),\\
    P_{4}&= s_{12}s_{45} + s_{23}(s_{34}-s_{56}-s_{345}) + 2 s_{34}^2 +s_{34} (-2 s_{56} +s_{61} +s_{123} -2 (s_{234}+s_{345})) \notag \\
    &\qquad-s_{56}s_{61} + s_{61}s_{123} + 2 s_{56}s_{234} - s_{123}s_{234} + s_{345} (s_{56} - s_{123} +s_{234}),
\end{align}
\begin{align}
    W_{260}&=\dfrac{-s_{12}s_{45}s_{234}+s_{34}s_{61}s_{123}+s_{345}(-s_{23}s_{56}+s_{123}s_{234})-\Delta_6}{-s_{12}s_{45}s_{234}+s_{34}s_{61}s_{123}+s_{345}(-s_{23}s_{56}+s_{123}s_{234})+\Delta_6}, \notag \\
    W_{i+260}&=\mathcal{T}^{i} (W_{260}), \qquad i=1,2,
\end{align}
\begin{align}
    W_{263}&=-\dfrac{P_5-(s_{12}\eps_ {1456} + s_{123} \eps_ {1256})}{P_5+(s_{12}\eps_ {1456} + s_{123} \eps_ {1256})}, 
    &W_{i+263}&=\mathcal{T}^{i} (W_{263}), &i&=1,\ldots,5, \\
    W_{269}&=-\dfrac{P_6-(s_{234} \eps_{6123}-s_{61} \eps_{1234})}{P_6+(s_{234} \eps_{6123}-s_{61} \eps_{1234})}, 
    &W_{i+269}&=\mathcal{T}^{i} (W_{269}), &i&=1,\ldots,5,
\end{align}
here the polynomials $P_5$ and $P_6$ are defined as follows
\begin{align}
    P_5&=s_{12} \left(s_{23} s_{56}-s_{61} s_{56}+s_{45} \left(s_{234}-s_{56}\right)\right) +s_{123} \left(-s_{34} s_{61}-s_{234} s_{345}+s_{56} \left(s_{61}+s_{345}\right)\right), \\
    P_6 &=s_{12} \left(s_{23} \left(s_{61}-s_{234}\right)+s_{45} s_{234}\right)+s_{123} \left(s_{234} s_{345}-s_{34} s_{61}\right) +s_{23} \left(s_{34} s_{61}-s_{56} s_{61}-s_{234} s_{345}\right).
\end{align}

Finally, we have 15 letters that transform non--trivially under the parity transformations. They are even under simultaneous changes of the sign of a square root $r_i$ and a pseudo--scalar $\eps_{ijkl}$, but odd under the sign change of just one of them.
\begin{align}
    W_{275}&=\dfrac{P_7-r_1\eps_{1234}}{P_7+r_1\eps_{1234}}, 
    &W_{i+275}&=\mathcal{T}^{i} (W_{275}), &i&=1,\ldots,5, \\
    W_{281}&=\dfrac{P_8-r_4\eps_{1234}}{P_8+r_4\eps_{1234}}, 
    &W_{i+281}&=\mathcal{T}^{i} (W_{281}), &i&=1,\ldots,5, \\
    W_{287}&=\dfrac{P_9-r_3\eps_{1245}}{P_9+r_3\eps_{1245}}, 
    &W_{i+287}&=\mathcal{T}^{i} (W_{287}), &i&=1,2,
\end{align}
where $P_7$, $P_8$ and $P_9$ are the following polynomials:
\begin{align}
    P_7&=s_{12}^2(s_{23}-s_{234})+s_{12} \left( -2s_{23} (s_{34} +s_{56}) +s_{34}(-2s_{56}+s_{123}) +s_{234}(s_{34}+s_{56}+s_{123})\right) \notag \\
    &\quad +(s_{34}-s_{56}) \left(s_{23}(s_{34}-s_{56}) +s_{123}(-s_{34}+s_{234}) \right),\\
    P_8&=s_{12} \left(s_{23}(2s_{56}-s_{123}+s_{234}) - s_{234}(s_{123}+s_{234}) \right)  -s_{123}(s_{34}-s_{234})(s_{123}+s_{234})  \notag \\
    &\quad+ 2s_{23}^2s_{56}+ s_{23} \left(s_{34}(2s_{56} +s_{123}-s_{234}) -s_{56}(s_{123}+s_{234})-2s_{123}s_{234} \right),\\
    P_9&=s_{12}s_{45} \left(2s_{23}+2s_{34}+2s_{56}+2s_{61}-s_{123} -4s_{234} -s_{345} \right) + (s_{123}+s_{345})\cdot \notag \\
   & \left(s_{23}(s_{34}-s_{56}-s_{345})-s_{61}(s_{34}-s_{56})-s_{123}(s_{34}+s_{61}-s_{234}) -s_{345}(s_{56}-s_{123}-s_{234})  \right).
\end{align}
Note that we used the subscript for $i=1,\ldots 289$ to list the alphabet with $D$-dimensional external momenta. A subset of $245$ letters that are independent for the four-dimensional external momenta are selected via the following relations:
\begin{align}
    \mathcal{T}^i &\left(\log{W_{191}}=-\log{W_{184}} - \log{W_{187}}-\log{W_{188}} \right), \quad i=0,1,2, \\
    \mathcal{T}^i &\left(\log{W_{214}}=\log{W_{184}} + \log{W_{185}}+\log{W_{187}}+\log{W_{188}}+\log{W_{210}} \right), \quad i=0,1,2, \\
    \mathcal{T}^i &\left(\log{W_{217}}=\log{W_{182}} - \log{W_{188}} + \log{W_{207}} \right), \quad i=0,1,2, \\
    \mathcal{T}^i &\left(\log{W_{221}}=-\log{W_{182}} + \log{W_{183}}+\log{W_{187}}+\log{W_{188}}-\log{W_{189}}+\log{W_{190}} \right. \notag \\
     &\left. \qquad \qquad \; +\log{W_{194}} -\log{W_{197}} +\log{W_{218}} \right), \quad i=0,1,2, \\
     \mathcal{T}^i &\left(\log{W_{248}}=\log{W_{182}} + \log{W_{185}} \right), \quad i=0,1,2, \\
     \mathcal{T}^i &\left(\log{W_{251}}=-\log{W_{194}} - \log{W_{195}}-\log{W_{218}} -\log{W_{219}} \right), \quad i=0,1, \\
     &\log{W_{253}}= \log{W_{182}} -\log{W_{183}}-\log{W_{187}}-\log{W_{188}}+\log{W_{189}}-\log{W_{190}} \notag \\
     &\qquad \qquad -\log{W_{194}}-\log{W_{196}}-\log{W_{218}}-\log{W_{220}} \\
     &\log{W_{254}}= \log{W_{182}} -\log{W_{183}}-\log{W_{186}}-\log{W_{187}}-\log{W_{188}}-\log{W_{190}} \notag \\
     &\qquad \qquad -\log{W_{194}}-\log{W_{218}}-\log{W_{229}} \\
     \mathcal{T}^i &\left(\log{W_{255}}=-\log{W_{182}} + \log{W_{183}} + \log{W_{188}} -\log{W_{189}} -\log{W_{195}} -\log{W_{219}} +\log{W_{224}}\right), \quad i=0,1, \\
     \mathcal{T}^i &\left(\log{W_{257}}=\log{W_{183}} - \log{W_{189}} - \log{W_{195}} -\log{W_{219}} -\log{W_{226}} \right), \quad i=0,1, \\
     &\log{W_{259}}= \log{W_{182}} -\log{W_{183}}+\log{W_{184}}+\log{W_{185}}+\log{W_{189}}-\log{W_{190}} \notag \\
     &\qquad \qquad -\log{W_{194}}-\log{W_{218}}-\log{W_{228}} \\
     \mathcal{T}^i &\left(\log{W_{262}}=\log{W_{183}} + \log{W_{190}} \right), \quad i=0,1, \\
     &\log{W_{261}}= -\log{W_{182}} - \log{W_{189}}, \\
     \mathcal{T}^i &\left(\log{W_{263}}=\log{W_{187}} - \log{W_{230}} \right), \quad i=0,\ldots,5, \\
     \mathcal{T}^i &\left(\log{W_{271}}=\log{W_{188}} + \log{W_{241}}\right), \quad i=0,1,2, \\
     \mathcal{T}^i &\left(\log{W_{274}}=-\log{W_{184}} - \log{W_{187}}- \log{W_{188}} + \log{W_{238}}\right), \quad i=0,1,2, \\
     \mathcal{T}^i &\left(\log{W_{279}}=-\log{W_{275}} - \log{W_{277}}\right), \quad i=0,1, \\
     \mathcal{T}^i &\left(\log{W_{287}}=-\log{W_{283}} - \log{W_{286}}\right), \quad i=0,1,2. 
\end{align}

Finally, the remaining 18 letters appearing beyond weight four are listed in equations~\eqref{eq:NewF4even} - \eqref{eq:NewF4odd}.

\section{Conversion Between the Previous 18 Letters in Ref.~\cite{Henn:2021cyv} and the New Alphabet}
\label{app:OldToNewF4}
In this section of the Appendix, we present the conversion table between the previous $18$ letters, associated with $F_4$, defined in Ref.~\cite{Henn:2021cyv} and the new alphabet $\mathbb A$.

In~\cite{Henn:2021cyv}, the old definition of letters also used the name ``$W$". To avoid confusion, in this work, we label the previous letters in~\cite{Henn:2021cyv}
with the notation $\tilde W$. In this notation, the letters $\tilde W_{188}\sim \tilde W_{205}$ in~\cite{Henn:2021cyv} are associated with $F_4$, and converted to the new alphabet $\mathbb A$ as
\begin{equation}
\begin{aligned}
& \log(\tilde W_{188}) = \log(W_{290}), \quad
\log(\tilde W_{189}) = \log(W_{291}), \quad
\log(\tilde W_{190}) = \log(W_{292}), \\
& \log(\tilde W_{191}) = \log(W_{293}) + \log(W_{296}) - \log(W_{299}), \quad
\log(\tilde W_{192}) = \log(W_{294}) + \log(W_{297}) - \log(W_{300}), \\
& \log(\tilde W_{193}) = \log(W_{295}) + \log(W_{298}) - \log(W_{301}), \quad
 \log(\tilde W_{194}) = \log(W_{293}) - \log(W_{296}) - \log(W_{299}), \\
& \log(\tilde W_{195}) = \log(W_{294}) - \log(W_{297}) - \log(W_{300}), \quad
 \log(\tilde W_{196}) = \log(W_{295}) - \log(W_{298}) - \log(W_{301}), \\
& \log(\tilde W_{197}) = \log(W_{293}) - \log(W_{296}) + \log(W_{299}), \quad
 \log(\tilde W_{198}) = \log(W_{294}) - \log(W_{297}) + \log(W_{300}), \\
& \log(\tilde W_{199}) = \log(W_{295}) - \log(W_{298}) + \log(W_{301}), \\
&
\log(\tilde W_{200}) = \log(W_{302}), \quad
\log(\tilde W_{201}) = \log(W_{303}), \quad
\log(\tilde W_{202}) = \log(W_{304}), \\
& \log(\tilde W_{203}) = \log(W_{305}), \quad
\log(\tilde W_{204}) = \log(W_{306}), \quad
\log(\tilde W_{205}) = \log(W_{307})\,.
\end{aligned}
\end{equation}

\bibliographystyle{JHEP}
\bibliography{main}

\end{document}